%% file: WFI2033_spec_z.tex
\documentclass[fleqn,usenatbib]{mnras}

\usepackage{newtxtext,newtxmath}

\usepackage[T1]{fontenc}
\usepackage{ae,aecompl}

\usepackage{graphicx}	
\usepackage{amsmath}	
\usepackage{amssymb}	
\usepackage{natbib}
\usepackage{footnote}
\usepackage[usenames]{color}
\usepackage{tabularx}
\usepackage{hhline}
\usepackage{array}
\usepackage{subfig}
\usepackage{url}
\usepackage{multirow}
\usepackage{booktabs}
\usepackage{hyperref}
\usepackage{xspace}

\usepackage{pdflscape}
\usepackage{longtable}
\usepackage{tablefootnote}
\usepackage{lineno}   
\usepackage{eso-pic} 

\newcommand{\cref}[1]{{\textbf{#1}}}
\newcommand{\rusu}{Rusu et al. (2019)}  
\newcommand{\wong}{Wong et al.  (2019)} 
\newcommand{\bonvin}{Bonvin et al. (2019)}   
\newcommand{\rusup}{Rusu et al. 2019}  
\newcommand{\wongp}{Wong et al.  2019} 
\newcommand{\bonvinp}{Bonvin et al. 2019}   

\def\HEofor{HE\,0435$-$1223}
\def\WFItwenty{WFI\,2033$-$4723}

\def\HOLI{H0LiCOW\xspace}
\def\kms{km\,s$^{-1}$}
\def\zs{1.662}
\def\zlens{0.6575}
\def\kmsMpc{\rm km\,s$^{-1}$\,Mpc$^{-1}$}
\title[Environment of \WFItwenty]{H0LiCOW\,X. Spectroscopic/imaging survey and galaxy-group identification around the strong gravitational lens system \WFItwenty}

\author[D. Sluse et al.] 
{D.~Sluse$^{1}$\thanks{dsluse@uliege.be}, 
C.~E.~Rusu$^{2, 3}$\thanks{Subaru Fellow},
C.~D.~Fassnacht$^{3}$,
A.~Sonnenfeld$^{4, 23}$, 
J.~Richard$^{11}$, 
\newauthor 
M.~W.~Auger$^{10}$, 
L.~Coccato$^{9}$, 
K.~C.~Wong$^{4, 5}$,
S.~H.~Suyu$^{6, 7, 8}$, 
T.~Treu$^{12}$,
A.~Agnello$^{9}$,
\newauthor 
S.~Birrer$^{12}$ ,
V.~Bonvin$^{13}$, 
T.~Collett$^{14}$,
F.~Courbin$^{13}$,
S.~Hilbert$^{15}$,
L.~V.~E.~Koopmans$^{16}$,
\newauthor 
O.~Tihhanova$^{13}$,
P.~J.~Marshall$^{17}$,
G.~Meylan$^{13}$, 
A.~J.~Shajib$^{12}$,
J.~Annis$^{18}$,
S.~Avila$^{19}$,
\newauthor 
E.~Bertin$^{20, 21}$,
D.~Brooks$^{22}$,
E.~Buckley-Geer$^{18}$,
D.~L.~Burke$^{17, 24}$,
A.~Carnero~Rosell$^{25, 26}$,
\newauthor 
M.~Carrasco~Kind$^{27, 28}$,
J.~Carretero$^{29}$,
F.~J.~Castander$^{30, 31}$,
L.~N.~da Costa$^{26, 32}$,
\newauthor 
J.~De~Vicente$^{25}$,
S.~Desai$^{33}$,
P.~Doel$^{22}$,
A.~E.~Evrard$^{34, 35}$,
B.~Flaugher$^{39}$,
J.~Frieman$^{18, 36}$,
\newauthor 
J.~Garc\'ia-Bellido$^{19}$,
D.~W.~Gerdes$^{34, 35}$,
D.~A.~Goldstein$^{37}$,
R.~A.~Gruendl$^{27, 28}$,
\newauthor 
J.~Gschwend$^{26, 32}$,
W.~G.~Hartley$^{22, 38}$,
D.~L.~Hollowood$^{39}$,
K.~Honscheid$^{40, 41}$,
D.~J.~James$^{42}$,
\newauthor 
A.~G.~Kim$^{43}$,
E.~Krause$^{44}$,
K.~Kuehn$^{45}$,
N.~Kuropatkin$^{18}$,
M.~Lima$^{46, 47}$,
H.~Lin$^{48}$,
\newauthor 
M.~A.~G.~Maia$^{26, 32}$,
J.~L.~Marshall$^{47}$,
P.~Melchior$^{48}$,
F.~Menanteau$^{27, 28}$,
R.~Miquel$^{49, 50}$,
\newauthor 
A.~A.~Plazas$^{48}$,
E.~Sanchez$^{25}$,
S.~Serrano$^{30, 31}$,
I.~Sevilla-Noarbe$^{25}$,
M.~Smith$^{51}$,
\newauthor 
M.~Soares-Santos$^{52}$,
F.~Sobreira$^{53, 26}$,
E.~Suchyta$^{54}$,
M.~E.~C.~Swanson$^{28}$,
G.~Tarle$^{35}$ \\
Affiliations appear at the end of the paper }

\date{Accepted XXX. Received YYY; in original form ZZZ}

\pubyear{2019}

\begin{document}
\label{firstpage}
\pagerange{\pageref{firstpage}--\pageref{lastpage}}
\maketitle


\begin{abstract}
	
{Galaxies and galaxy groups located along the line of sight towards gravitationally lensed quasars produce high-order perturbations of the gravitational potential at the lens position. When these perturbation are too large, they can induce a systematic error on $H_0$ of a few-percent if the lens system is used for cosmological inference and the perturbers are not explicitly accounted for in the lens model. In this work, we present a detailed characterization of the environment of the lens system \WFItwenty\,($z_{\rm src} =\,$\zs, $z_{\rm lens}=\,$\zlens), one of the core targets of the \HOLI project for which we present cosmological inferences in a companion paper (\rusup). We use the Gemini and ESO-Very Large telescopes to measure the spectroscopic redshifts of the brightest galaxies towards the lens, and use the ESO-MUSE integral field spectrograph to measure the velocity-dispersion of the lens ($\sigma_{\rm {los}}= 250^{+15}_{-21}$ \, \kms) and of several nearby galaxies. In addition, we measure photometric redshifts and stellar masses of all galaxies down to $i < 23$ mag, mainly based on Dark Energy Survey imaging (DR1). Our new catalog, complemented with literature data, more than doubles the number of known galaxy spectroscopic redshifts in the direct vicinity of the lens, expanding to 116 (64) the number of spectroscopic redshifts for galaxies separated by less than 3\arcmin\,(2\arcmin) from the lens. Using the flexion-shift as a measure of the amplitude of the gravitational perturbation, we identify  2 galaxy groups  and 3 galaxies that require specific attention in the lens models. The ESO MUSE data enable us to measure the velocity-dispersions of three of these galaxies. These results are essential for the cosmological inference analysis presented in \rusu.}

\end{abstract}

\begin{keywords}
gravitational lensing: strong -- quasars: individual: \WFItwenty -- galaxies: groups: general
\end{keywords}

\AddToShipoutPictureBG*{%
	\AtPageUpperLeft{%
		\hspace{0.75\paperwidth}%
		\raisebox{-3.5\baselineskip}{%
			\makebox[0pt][l]{\textnormal{DES-2019-0432}}
}}}%

\AddToShipoutPictureBG*{%
	\AtPageUpperLeft{%
		\hspace{0.75\paperwidth}%
		\raisebox{-4.5\baselineskip}{%
			\makebox[0pt][l]{\textnormal{FERMILAB-PUB-19-168-AE}}
}}}%



\section{Introduction} 
\label{sec:intro}

The spectroscopic identification of the galaxies located in the environment or along the line-of-sight towards a gravitational lens, is one of the important tasks to carry out for deriving an accurate time-delay distance. This is particularly relevant because the lensing cross section is larger for galaxies residing in rich environment \citep{Fassnacht2011, Wong2018b}. Depending on their redshift and projected distance from the main lens, galaxies or galaxy groups, may significantly perturb the light bending produced by gravitational lensing. The amplitude of the perturbation on the lensed images is larger when the perturber is located in the foreground of the lens, and is maximum at the lens redshift \citep{McCully2017}. The influence on the lensed images also depends on the projected distance of the perturber to the lens. When sufficiently distant in projection to a lens system, galaxy groups (or clusters) produce a uniform convergence at the lens position. This effect can be accounted for in the time-delay distance estimate on a statistical basis, following a methodology similar to the one developed in \cite{Rusu2017}. When closer in projection to the lens, galaxies or galaxy groups produce higher-order perturbations to the gravitational potential, and therefore must be explicitly included in the lens model; otherwise these perturbations introduce an unknown systematic error. The shift in lensed image positions derived by comparing models with or without the perturber (i.e. the so called flexion shift), may be used as a criterion to identify objects that need to be included explicitly in the lens model \citep{McCully2017}. For these reasons, it is crucial to obtain spectroscopic and photometric redshifts of the brightest galaxies observable in the field of view (FOV) of a lens system.

The  \HOLI ($H_0$ Lenses in COSMOGRAIL's Wellspring) program has been initiated with the aim of measuring the Hubble constant $H_0$ with better than $3.5\%$ accuracy from a small sample of gravitationally lensed quasars with a diversity of observational properties \citep{Suyu2017}. To reach this goal, the program combines several ingredients: it gathers high-quality data (\emph{Hubble Space Telescope} (HST) imaging, deep images of the FOV, medium resolution spectroscopy of the lens and of nearby galaxies) for each scrutinized system \citep{Suyu2017}, a few-percent accuracy measurement of the time delays \citep{Bonvin2016}, and advanced Bayesian lens-modelling techniques \citep{Suyu2010b, Suyu2012b,  Birrer2015a, Birrer2018}. An important aspect of the \HOLI methodology is that the inferred value of the cosmological parameters (encoded into the so-called time-delay distance) remains blinded until publication. The results are unblinded only when the collaboration considers that all necessary measurements, modelling and tests have been performed, and then published ``as is''. 

\WFItwenty\,is part of the \HOLI main sample of time-delay lenses. It is a quadruply lensed quasar at redshift $z_{\rm src} =\,$\zs\, lensed by an elliptical galaxy at $z_{\rm lens}=\,$\zlens$\,\pm\,0.001$ \citep[][ this paper]{Morgan2004, Eigenbrod2006b, Sluse2012b}. The minimum and maximum image separation are respectively of $\Delta \theta_{\rm{min}} \sim 0.8\arcsec$\, and $\Delta \theta_{\rm{max}} \sim 2.5\arcsec$, such that the two brightest images are only barely spatially resolved with ground-based and natural-seeing data, but the two other images are easily photometrically monitored with a 1-m class telescope. Time-delay measurements for the various combinations of image pairs are presented in \bonvin. When this system was selected to be part of the \HOLI sample, the line-of-sight towards the lens was yet to be characterized. An important step forward in the characterization of the lens environment through spectroscopy has been carried out by \cite{Momcheva2015} and \cite{Wilson2016}. In particular, \cite{Wilson2016} have confirmed that the lens is part of a massive galaxy group as first suspected by \cite{Morgan2004}. Because a proper characterization of the lens environment is crucial to control the systematic errors on $H_0$, we have carried out a deeper spectroscopic survey of the FOV of \WFItwenty, derived photometric redshifts for the faintest field galaxies, and estimated their stellar masses. Owing to the ESO-MUSE instrument \citep{Bacon2010}, we have been able to carry over a more exhaustive characterization of the galaxies closest to the lens in projection, measuring their redshifts as well as the velocity-dispersions of the lens and of its brightest neighbours. The description and analysis of those new observations, which double the number of spectroscopic measurements for the nearest (in projection) field galaxies, are the main purpose of the this paper. They are used to identify and get a proxy on the mass of the main perturbers of the lens potential that need to be explicitly included in the lens-modelling for cosmological inference (\rusup). A joined cosmological inference based on all the lensed systems measured so far by \HOLI is presented in \wong. 

The paper is structured as follows. We present an overview of the data sets used and of the data reduction process in Sect.~\ref{sec:data}. The techniques employed to measure the photometric and spectroscopic redshifts (hereafter photo-z and spec-z respectively) and stellar masses are presented in Sect.~\ref{sec:redshifts}. The methodology used to identify galaxy groups and a list of the groups we identified are described in Sect.~\ref{sec:groups}. Section~\ref{sec:model} quantifies the impact of individual galaxies and galaxy groups on the model. We use the \emph{flexion shift} to flag the systems that require explicit inclusion in the multi-plane lens models presented for this system by \rusu. We further measure their velocity-dispersions in Sect.~\ref{sec:vdisp}, as this information is included in the lens-modelling presented by \rusu. In addition, we also measure the velocity-dispersion of the lensing galaxy, which is instrumental in reducing the impact of the mass-sheet degeneracy on the lens models. Finally, Sect.~\ref{sec:conclude} summarizes our main results. In this work, with the exception of the target selection that was based on $R-$band magnitude in the Vega system, photometric information comes from multicolor imaging and uses the AB photometric system. For convenience, group radii and masses reported in this work assume a flat $\Lambda$CDM cosmology with $H_0 = 70 $\,\kmsMpc, and $\Omega_{\rm {m}} = 0.3$. We stress that this choice has no impact on the group identification as this does not depend on a specific choice of cosmological parameters. 


\section{Data} 
\label{sec:data}

Our data set combines multi-object and integral field spectroscopy obtained with Gemini-South and ESO-Paranal observatories, and multi-band/deep imaging obtained with the Spitzer Space Telescope and the Blanco Telescope, including data from the Dark Energy Survey (DES\footnote{\url{https://www.darkenergysurvey.org}}). The goals of the spectroscopic observations are to measure accurate redshifts and identify galaxy groups which need to be explicitly accounted for in the lens model; to measure velocity-dispersions for the massive individual galaxies that are close enough to also require inclusion in the lens model; and to calibrate the photometric redshifts extracted for the galaxies in the imaging data without available spectroscopy. The multi-band imaging data complement the spectroscopy, while allowing the measurement of photometric redshifts and stellar masses of galaxies up to a fainter magnitude limit (our setup yields a typical depth of $i \sim$ 23\,mag). Those data are also crucial for the cosmographic analysis as they are instrumental to the estimation of the distribution of convergence at the lens position \citep[see][\rusup]{Rusu2017}. A summary of the data sets is provided in Table~\ref{tab:data}.

\subsection{Imaging}
\label{subsec:obs_imaging}

Homogeneous, multi-band, large FOV imaging observations are needed in order to achieve a more complete characterization of the environment and the line of sight (LOS) of \WFItwenty\ than what is possible through targeted spectroscopy. We base our analysis mainly on $grizY$-band DES data included in the Data Release 1 \citep{abbott18} and obtained during 2013 September and 2015 September (2014 September - 2015 October for the $z-$band). We supplement this with proprietary deep $u-$band data observed on 2015 July 21, 22 (PI. C.\,E. Rusu) with the Dark Energy Survey Camera \citep{flaugher15} on the Blanco Telescope; VLT/HAWK-I \citep{pirard04,kissler08} near-infrared data (PI. C.\,D. Fassnacht, program ID 090.A-0531(A)) observed on 2012 October 12; and with archival IRAC \citep{fazio04} infrared data from the Spitzer Space Telescope (PI. C. S. Kochanek, program ID 20451), observed on 2005 October 20 and 2006 June 4. The characteristics of our data are described in Table~\ref{tab:data}. We also have WFC3 F160W {\emph {HST}} imaging data (PI. S.\,H. Suyu, Program ID 12889) from 2013 April 3 and 4, which is presented in more detail by \citet{Suyu2017} and was only used in this work to check the quality of the star-galaxy classification (see Section \ref{sec:photozmstar} below). 

\subsection{Spectroscopy}
\label{subsec:spectro}

 The use of multi-object spectroscopy is optimal to identify group(s) or cluster(s) of galaxies with projected distances of several arcmin from the lens (i.e. typically a few virial radii for groups at $z > 0.1$). In this work, we used the MXU capabilities (multi-object spectroscopy mode with exchangeable laser-cut masks) of the FORS2 instrument \citep{Appenzeller1998} mounted at the Cassegrain focus of the UT1 (Antu) telescope (PID: 091.A-0642(A), PI: D.\,Sluse), and the multi-object spectroscopy mode of the Gemini Multi-Object Spectrographs \citep[GMOS;][]{Hook2004} at the Gemini-South telescope (PID: GS-2013A-Q-2, PI: T. Treu). The instrumental setup and target selection strategy is similar to the one we used for the lens system \HEofor\, and we refer the reader to \cite{Sluse2017} for details on the latter. In brief, we used 6 masks and the  GRIS300V grism\,+\,GG435 blocking filter for the FORS2 data, and 4 masks and the R400 grating with GG455 filter for the GMOS observations. The 2 instruments allow us to put slits on targets located up to typically 2\arcmin\, from the lens, and pack approximately 35 long-slits (6\arcsec\, length) per mask. With a 40 minutes exposure time per mask (yielding 1h execution time with overheads), we can measure redshifts of galaxies up to magnitudes $I\sim 21.5$. This setup maximizes the number of observable targets and ensures a large wavelength coverage (typically 4500-8700\AA) to ease redshift detectability. During the observations, the seeing was always lower than 0.9\arcsec, and airmass $1 < \sec(z) < 2$. The FORS2 observations were carried out in service mode between 2013-05-31 and 2014-09-13, while GMOS data were obtained in visitor mode during the nights of 2013-06-03 and 2013-06-06. 

The exceptional capabilities of the ESO-MUSE Integral Field Spectrograph, mounted at the Nasmyth B focus of Yepun (ESO-VLT UT4 telescope), offer a natural complement to the multi-object data. Owing to its wide FOV of $1\arcmin\times 1\arcmin$, and a $0.2\arcsec \times 0.2\arcsec$ spatial sampling, it allows one to obtain 90000 simultaneous spectra covering almost the whole optical range (4800-9350\,\AA) with a resolving power $R\sim 1800-3600$ \citep[i.e. 2.5\,\AA~spectral resolution; ][]{Richard2017}. It is therefore perfectly designed to characterize the lens environment on small scales, allowing the measurement of the redshift of the nearest perturbers,  and of the velocity-dispersion of the brightest galaxies (including the lensing galaxy). Our observing strategy consist of placing the lensing galaxy close to the centre of the field and obtaining 4 exposures of 600\,s, each rotated by 90 degrees with respect to the previous one, and offset by a few spaxels (spatial pixels). The four exposures of 600\,s are combined into a single data cube of 2400\,s during the data reduction. A first ensemble of 3 combined data cubes has been obtained as part of the Science Verification (SV) programme 60.A-9306(A), on 2014-06-19 and 2014-08-24, allowing us to reach a depth of $I \sim $ 25 mag (continuum emission, 3$\sigma$). A second ensemble of 6 data cubes (Wide Field mode) has been obtained in Service mode on 2016-05-24, 2016-06-29, 2016-07-18, 2016-07-19, 2016-07-20, under programme 097.A-0454(A) (PI: D. Sluse; hereafter P97). Conditions are optimal (i.e. clear sky, seeing better than  $0.8\arcsec$) only for a fraction of the P97 data. According to the grading scheme established by ESO, two data sets are attributed a grade A (conditions similar to SV data, fulfilled), one a grade B (marginally out of specification), and three a grade C (out of specification). The P97 data are obtained under high moon fraction, and are therefore less deep than the SV data, with depth between $I \in [21.3, 24.9 ]$ mag. 

\begin{table*}[!h]
	\caption{{\bf Overview of the imaging and spectroscopic data set}. For spectroscopy, the columns list respectively the instrument used, the number of masks (except for the data obtained with the ESO-MUSE integral field spectrograph), the total number of objects targeted, the approximate resolving power $R$ of the instrument at central wavelength, the typical wavelength range covered by the spectra (spectra do not always cover the full wavelength range, depending on the exact object location in the field), and the exposure time per mask, or for the full data set in case of ESO-MUSE data. Note that the \# of spectra includes duplicated objects. For imaging, the columns list the magnitude depth, filter name, seeing and exposure time of the data sets used. }
	\label{tab:data}
	\begin{tabular}{lccccc}
		\toprule
		Instrument: & \# of  & \# of & $R$ & $\lambda_1-\lambda_2$  & Exp  \\
		\small{Spectroscopy} & Masks & spectra &   & (\AA)  & (s) \\ 

		\midrule
		FORS2  & 6 & 236 & 440 &4500-9200 & 2$\times$1330\\
		GMOS  & 4 & 130 & 1000 & 4400-8200& 4$\times$660 \\	
		MUSE$^\dagger$  & NA &  20 & 1800-3600 & 4800-9400 & 9$\times$4$\times$600 \\
		\hline
		Imaging$^\ddagger$ & depth$^\star$ & filter & scale & seeing & Exp \\
		& [mag] &         & [$\arcsec$] & [$\arcsec$] & (s) \\ 
		\hline
		HAWK-I & $21.5\pm0.1$ & $J$ & 0.1064 & 0.71 & $7\times67.5$ \\	
		HAWK-I & $20.86\pm0.08$ & $H$ & 0.1064 & 0.71 & $3\times60$ \\	
		HAWK-I & $20.76\pm0.04$ & $Ks$ & 0.1064 & 0.60 & $3\times60$ \\	
		DECam  & $25.17\pm0.06$ & $u$ & 0.2625 & 1.16 & $65\times500$ \\	
		DES & $24.25\pm0.05$ & $g$ & 0.2625 & 1.21 & $5\times90$  \\
		DES & $23.8\pm0.1$ & $r$ & 0.2625 & 0.97 & $5\times90$  \\
		DES & $23.13\pm0.08$ & $i$ & 0.2625 & 0.81 & $6\times90$  \\
		DES & $22.9\pm0.5$ & $z$ & 0.2625 & 1.16 & $4\times90$  \\
		DES & $21.4\pm0.2$ & $Y$ & 0.2625 & 0.92 & $7\times45$  \\
		IRAC & $24.6\pm0.3$ & 3.6 & 0.600 & - & $72\times30$ \\
		IRAC & $24.0\pm0.2$ & 4.5 & 0.600 & - & $72\times30$  \\
		IRAC & $22.3\pm0.3$ & 5.7 & 0.600 & - & $72\times30$  \\
		IRAC & $22.1\pm0.3$ & 7.9 & 0.600 & - & $72\times30$  \\
		WFC3 & $26.4\pm0.1$ & F160W & $0.08$ & - & 26257 \\		
		\bottomrule
	\end{tabular}
	{\small \\
		{\bf Notes:} $\dagger$ Only 4/9 data sets were obtained within requested observing conditions (graded A by ESO). The others were graded B (1/9) or C (4/9), which means that the seeing was not stable during an observation and/or moon was too close, yielding a high sky level. \\ 
		$\ddagger$ The number of exposures for DES data denotes the maximum number of overlaps, as the coverage is not uniform. The pixel scale and exposure time reported for WFC3 characterize the final frame obtained after combining dithered exposures with \texttt{DrizzlePac}\footnotemark\\ 
		 $^\star$ We measure $5\sigma$ detection limits as $m_\mathrm{lim} = \mathrm{ZP} - 2.5 \log\left(5 \sqrt{N_\mathrm{pix}}\sigma_\mathrm{sky}\right)$, where ZP is the magnitude zero-point, $N_\mathrm{pix}$ is the number of pixels in a circle with radius 2\arcsec, and $\sigma_\mathrm{sky}$ is the sky-background noise variation. We derive the uncertainty as the standard deviation of the values in 10 empty regions across the frame. }
	
\end{table*}

\footnotetext{\texttt{DrizzlePac} is a product of the Space Telescope Science Institute, which is operated by AURA for NASA. }  

\subsection{Spectroscopy data reduction}
\label{subsec:reduc}

We carried out data reduction of the FORS2 and Gemini multi-object spectroscopy data following the same prescriptions as \cite{Sluse2017}. The reduction cascade includes the standard steps of spectroscopic data reduction. They are implemented within the  ESO \texttt{reflex} environment \citep{Freudling2013} and FORS2 pipeline version 2.2 for FORS data, and through the {\it gemini-gmos} IRAF\footnote{IRAF is distributed by the National Optical Astronomy Observatories, which are operated by the Association of Universities for Research in Astronomy, Inc., under cooperative agreement with the National Science Foundation.} subpackage for GMOS data. Of particular relevance for this work is the accuracy at which the wavelength calibration has been performed. For FORS2 data, we used a polynomial of degree $n=5$, which yielded residuals distributed around 0, a RMS of typically 0.2 pixels = 0.66\,\AA\,at all wavelengths and a model accuracy estimated by matching the wavelength solution to the sky lines, to 0.25\,\AA. Comparison of spectra obtained with different instruments confirms the accuracy of the wavelength calibration (See Appendix~\ref{Appendix:redshift}). 

The MUSE data reduction has been carried out using the MUSE reduction pipeline version 2.0.1 \citep{Weilbacher2012, Weilbacher2015}. In particular, the standard steps of bias and flat-fielding corrections, wavelength solution, illumination correction, and flux calibration were made for each of the individual exposures with the default parameters of the pipeline. A variance data cube is associated to each data cube produced by the pipeline. It propagates the errors all along the pipeline reduction chain. While the SV data, obtained during dark observing conditions, are little affected by sky subtraction residuals, this is not the case with the P97 mode data. The latter have been post-processed using the Zurich Atmospheric Purge tool \citep[ZAP; ][]{Soto2016} that improves the sky subtraction by constructing a sky model using principal components analysis. For each data subset, a combined data cube, sampled on a grid of $0.2\arcsec \times 0.2\arcsec \times 1.25$\,\AA, is reconstructed. For the SV data, we combine the three individual data sets, yielding a total exposure time of 7200\,s and a median seeing of 1\arcsec. For the P97 data, we tested different combinations of data cubes, minimizing the seeing, amplitude of sky residuals, and optimizing the signal-to-noise ratio (SNR). We find that optimizing the SNR is essential for performing reliable velocity-dispersion measurements of the galaxies. The final datacube for P97 combines twelve exposures, for a total exposure time of 10800\,s.

We note that for FORS2 data, we sometimes included 2 objects in a slit to maximize the number of observed targets. For that reason, we perform the extraction by fitting a sum of 1-D Gaussian profile on each wavelength bin of the rectified 2-D spectrum (with n=[1,2] depending of the number of objects in the slit). The extraction is performed on individual exposures of each spectrum, and final 1-D spectrum is the result of the coaddition of the wavelength-calibrated extracted spectra of the same target.

\section{Redshifts and stellar masses } 
\label{sec:redshifts}

\subsection{Photometric redshifts and stellar masses}
\label{sec:photozmstar}

Here we give a brief description of our technique to measure photometric redshifts and stellar masses, which follows the technique described in \citet{Rusu2017}. The analysis of the resulting data for estimating the external convergence that is necessary for the cosmological inference will be presented by \rusu.

While the  DES and DECam image mosaics cover a very large FOV, the {\emph{HST}} data cover only the inner $\sim2.2\arcmin \times2.6\arcmin$ region. In addition, the IRAC and HAWK-I data cover just a few arcminutes around \WFItwenty; this is not enough to map the whole area where there is spectroscopic data, but is enough to map the $2\arcmin$ radius around the lensing system, where structure in the environment and along the LOS has the greatest impact on the lensing model \citep{Collett2013}. 

We downloaded cutouts covering $4\arcmin\times4\arcmin$ around \WFItwenty~using the DES cutout service\footnote{\url{https://des.ncsa.illinois.edu/easyweb/cutouts}}. These consist of $grizY$-band individual exposures that were processed by the DES pipeline \citep{morganson18} to remove the instrumental signature, including bias subtraction, flat-fielding, sky subtraction, artifact masking, and astrometric/photometric calibration. However, at the time when we performed the analysis, master coadded frames were not available. We therefore used \texttt{Scamp} \citep{Bertin2006} to ensure an accurate image registration, and performed image coaddition in each band with \texttt{Swarp} \citep{Bertin2002}. We followed similar steps to reduce the DECam $u-$band data (the same instrument used by DES), except that we could not achieve a viable photometric calibration, despite that the observing conditions seemed photometric (but the presence of thin cirrus cannot be excluded). 

We reduced the HAWK-I data using the recommended reduction pipeline\footnote{\url{https://www.eso.org/sci/software/gasgano.html}}, in conjunction with \texttt{Scamp} and \texttt{Swarp}, resampling onto the DES pixel scale, and we calibrated the absolute photometry using bright but unsaturated stars from 2MASS \citep{skrutskie06}. In order to enable the measurement of accurate colors between the different filters, we homogenized the shape and size of the point spread function (PSF) by applying suitable convolution kernels. These kernels were computed between two-dimensional Moffat profiles \citep{Moffat1969}  fitted in each band to scaled and stacked bright stars inside the FOV. The resulting PSF Full Width at Half Maximum (FWHM) was $\sim1.2\arcsec$. 

Our technique to perform object detection and photometric measurements follows that of \citet{Erben2013}. For each of the $ugrizYJHKs$ bands, \texttt{SExtractor} \citep{Bertin1996} is run in dual-image mode, where the detection image is the sum of the deepest, best-seeing DES images ($r$ and $i$, although we also performed detections in the $i-$band image only), and the  measurement images are the PSF-matched images in each of the filters. An additional run performs measurements in the original (i.e. not convolved) $i-$band image. This last run is performed to obtain total magnitudes (\texttt{SExtractor} quantity MAG\_AUTO), whereas the previous runs yield accurate colours based on isophotal magnitudes (MAG\_ISO). As our resampling and convolutions can produce large noise correlation, which may significantly underestimate the photometric uncertainties measured with \texttt{\texttt{SExtractor}}, we use the technique described in \citet{Labbe2003}, \citet{Gawiser2006} and \citet{Quadri2007} to correct for this effect. Finally, we downloaded reduced and photometrically calibrated IRAC data, and we used \texttt{T-PHOT} \citep{Merlin2015} to measure magnitudes matched to the apertures in the DES data, given the much larger pixel scale of the IRAC data and the broader PSF.  

\begin{figure}
\includegraphics[width=\columnwidth]{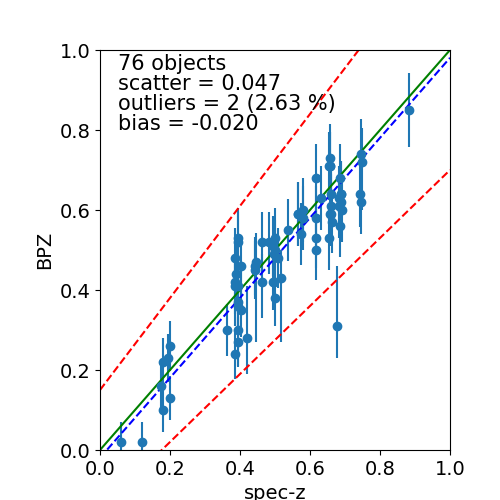}
\caption{Comparison of spectroscopic and photometric (\texttt{BPZ}) redshifts for galaxies with robust spectroscopic redshifts within the $120\arcsec$ radius around the lensing system, based on $ugrizYJHK$ photometry. The blue dashed line represents the best-fit offset, and the green solid line the perfect equality between the two redshift estimates. We define the outliers as data located outside the red dashed line marking $|z_\mathrm{spec} - z_\mathrm{phot}|/(1+z_\mathrm{spec})>0.15$. Error bars refer to $1\sigma$ uncertainties.
\label{fig:specz}}
\end{figure}

We adopt the galaxy-star classification of \citet{Hildebrandt2012}. Objects with $i < 21$ and with size smaller than the PSF are classified as stars. In the range $21< i <23$, an object is defined as a star if its size is smaller than the PSF and in addition if $\chi^2_\mathrm{star} < 2.0 \chi^2_\mathrm{gal}$, where $\chi^2$ is the best-fitting goodness-of-fit $\chi^2$ using galaxy and stellar templates. We use both \texttt{BPZ} \citep{Benitez2000} and \texttt{EAzY} \citep{Brammer2008} to measure photo-zs for the resulting galaxies. Similar to \citet[][]{Hildebrandt2010}, we find that the use of currently available mid-IR templates degrade rather than improve the quality of the inferred redshifts. We therefore ignore the IRAC data when estimating redshifts. While the $u-$band data were observed in non-photometric conditions, we solved for its zero point in a separate run with \texttt{BPZ} by minimizing the difference between photo-zs and spec-zs where available. Figure \ref{fig:specz} shows a comparison of the photo-zs and spec-zs, when the latter exist and are reliable. We also compared the photo-zs estimated with \texttt{BPZ} and \texttt{EAzY}. They agree well, with an average scatter of 0.06 and an average outlier fraction (i.e. objects with $|\Delta z| / (1+z) > 0.15$) of 11\% down to the magnitude limit of $i<23$ mag. 

Finally, since stellar masses are not direct output of \texttt{BPZ} and \texttt{EAzY}, we estimated stellar masses with \texttt{Le PHARE} \citep{Arnouts2002, Ilbert2010}, using galaxy templates based on the stellar population synthesis package of \citet{Bruzual2003} with a \citet{Chabrier2003} initial mass function (IMF). The stellar mass estimates are performed fixing the redshift to the best fitted photo-z. We report the photometry of the $i<23$ mag galaxies within $2\arcmin$ of \WFItwenty~ in Table \ref{tab:phot}, and the corresponding redshifts and stellar masses in Table \ref{tab:zmstar}. Those tables are also available in electronic form\footnote{\url{http://www.h0licow.org}}. 

In the above, we addressed the galaxies within $4\arcmin\times4\arcmin$ of \WFItwenty, where our data provides uniform coverage. For the surrounding FOV of up to $30\arcmin$ away, we rely on DES data to perform galaxy/star separation and measure photo-zs and stellar masses in a similar way. However, instead of performing our own measurements, we rely on total magnitudes provided by the DES pipeline in the form of the $\mathrm{Y3A1\_COADD\_OBJECT\_SUMMARY}$ table retrieved with \texttt{easyaccess} \citep{CarrascoKind2018}. This results in an increased fraction of photo-z outliers, from  $\sim3\%$ to $\sim14\%$. We make no effort to improve the extracted colors, as our only use of the resulting quantities is to explore the completeness of our spectroscopic redshifts (see Section \ref{subsec:completeness}). 

\subsection{Spectroscopic redshifts}
\label{subsec:specz}

We followed the methodology of \cite{Sluse2017} for the redshift measurements. Each combined 1-D spectrum of an object{\footnote{If an object was observed in several masks, redshift measurements were performed independently to avoid introducing biases due to uncertainties in wavelength calibration and/or differences of wavelength coverage.} is cross-correlated with a set of galactic (Elliptical, Sb, only galactic emission lines, quasar) and stellar ($G$, $O$, $M1$, $M8$, $A$ spectral types, composite of multiple spectral types) templates using the \texttt{xcsao} task, part of the \texttt{rvsao} IRAF package (version 2.8.0). Sky regions known to be contaminated by telluric absorption, and/or where sky subtraction is not satisfactory, are masked out. Redshift guesses are derived visually, and refined using the interactive mode of  \texttt{rvsao}. The redshift from the template providing the highest cross-correlation peak is considered as our final redshift measurement. A flag 0 (secure) / 1 (tentative) / 2 (insecure) is then attached to the spectrum based on the quality of the cross-correlation, signal-to-noise and number of emission/absorption lines detected. The uncertainty on the redshift derived with \texttt{xcsao} depends only on the width and peak of the cross-correlation. This error appears to be representative of the statistical uncertainty affecting our measurements, but is smaller than the systematic error as derived by comparing our spectra to literature data (See Appendix~\ref{Appendix:redshift}). Unless explicitly stated, the statistical error is used throughout this analysis. It is also the error reported in the final catalog. 

The galaxies detected in the MUSE FOV have been identified automatically using the {\texttt{MUSELET}} tool, part of the MPDAF package \citep{Piqueras2017}, applied on the combined Science Verification data cubes. Because of the almost dark conditions during the observations, those data allowed us to reach a $3 \sigma$ magnitude limit $AB = 25.3$ for a point source, i.e. more than 1 mag deeper than any combination of P97 data cubes. The \texttt{MUSELET}} tool performs an automatic detection of emission-line features in data cubes by flagging pixels that deviate from the noise \citep[see Sect. 2.2.1 of][for  a detailed description]{Drake2017}. A guess redshift is automatically derived, associating the observed features to brightest multiplets of emission-lines detected in galaxy spectra, or to Ly$\alpha$ emission if only one line is detected. The detection of emission in multiple consecutive pixels (along the spectral direction) is used to identify spurious line emission. We visualised the spectra of all the automatically-identified objects to flag obvious artefacts (e.g. sky reduction artefacts that concentrate close to the edge of the FOV). Finally, we compared the catalog of {\texttt{MUSELET}} targets to a catalog of objects detected by running \texttt{SExtractor} on the median data cube (i.e. median along the wavelength direction). This allows the identification of  objects that lack emission-lines. For all the targets we remeasured the redshift using \texttt{rvsao}, following the methodology described above. 

\begin{figure*}
	\includegraphics[scale=0.6]{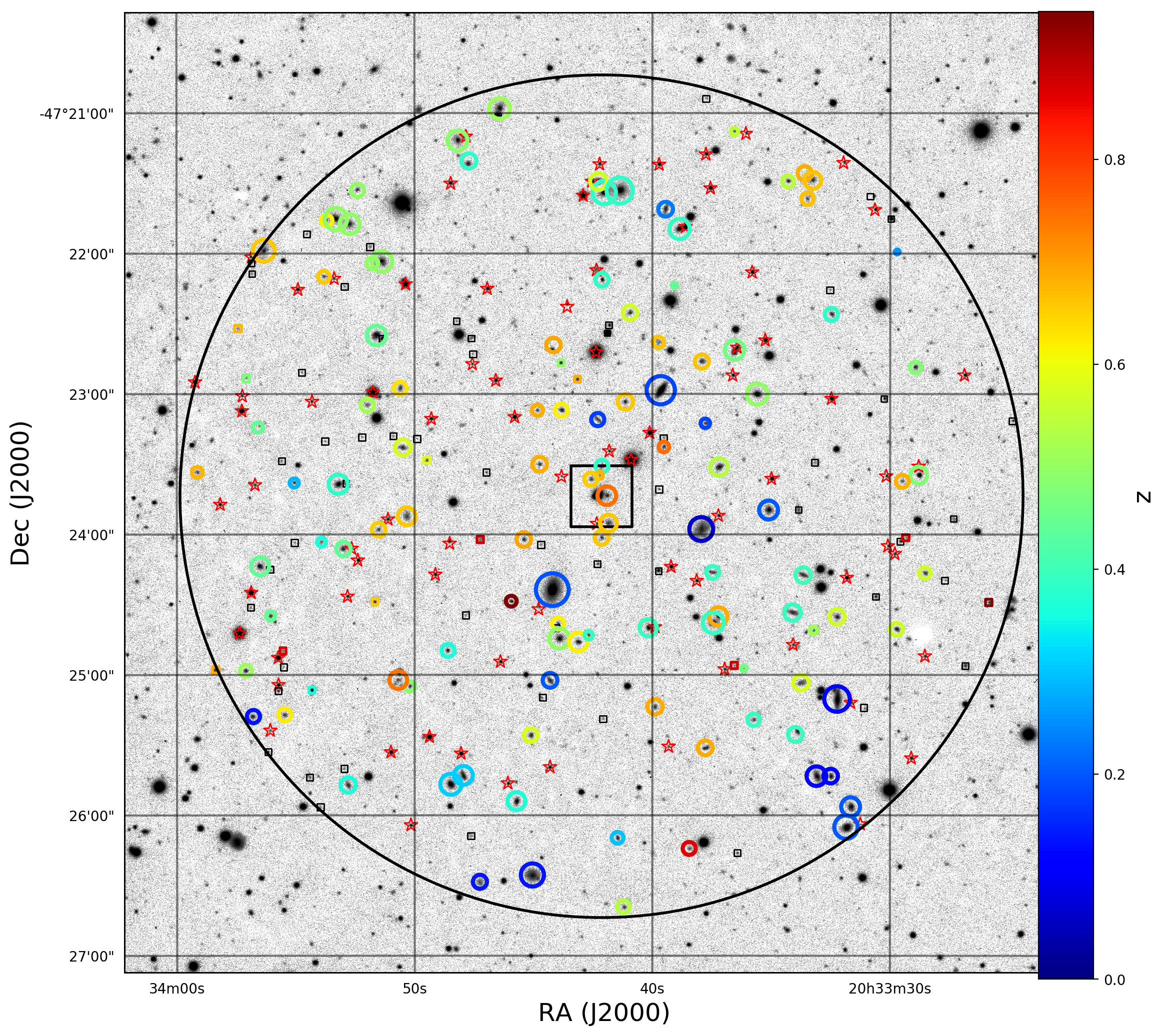}
	\caption{Overview of the spectroscopic redshifts obtained from our new and literature data in a FOV of $\sim$ 6\arcmin\,$\times$6\arcmin\, around \WFItwenty~(the black circle delimits a 180\arcsec\, radius FOV around the lens). Spectroscopically identified stars are marked with a red "Star" symbol, while galaxies are marked with a circle whose size scales with its $i$-band magnitude (largest colored circle correspond to $i \sim $18.6\,mag, smallest to $i\sim $23.9\,mag), and color indicates the redshift (right color bar). Galaxies that have been targeted but for which no spec-z could be retrieved are shown as open black squares, those with a tentative redshift (zQF = 1, see Table~\ref{tab:catalog}) with a colored square (right color bar). The background frame shows an archival FORS1 R-band combined frame (Prog. ID: 074.A-0563(A)) of 300\,s effective exposure time. A zoom on the central region is displayed in Fig.~\ref{fig:zoom}}
	\label{fig:field}
\end{figure*}

The last step consists in merging the various spectroscopic catalogs into a single one. For each spectrum, an approximate astrometric calibration is deduced based on information recorded in the header of the raw frame. For MXU data, only the position at the centre of the slit is recorded, such that we applied an additional correction based on the object position within the slit and orientation of the laser-cut mask on the sky. Because of the uncertainty of a few arcseconds on the absolute astrometric calibration of the various instruments, and of additional random uncertainties associated with spectral extraction, the astrometric positions between catalogs gathered with different instruments differ by up to 3\arcsec. Since there was a substantial number (i.e. $>$ 10) of objects in common between pairs of catalogs, we can cross-match catalogs to derive the median astrometric offset (in RA-DEC) ranging from 1.6\arcsec\, to 3.2\arcsec\, depending of the catalogs considered. Once all the catalogs are virtually matched to the same astrometric system, a new (more robust) cross-correlation can be performed, allowing us to identify duplicates and possible errors in redshift measurements. Objects present in multiple spectroscopic catalogs are found to have compatible redshifts. Instead of combining the multiple measurements, we have decided to keep only the entry with the lowest redshift uncertainty. The final merged catalog as well as the extracted spectra for the GMOS, FORS2 and MUSE data will be available upon acceptance of the paper in electronic form\footnote{\url{www.h0licow.org}}. The first 5 lines of the catalog are displayed in Table~\ref{tab:catalog}. Fig.~\ref{fig:field} provides an overview of the targets for which spectroscopic information has been gathered within 180\arcsec\, from \WFItwenty. 

The comparison between multiple data sets also provides a good way to flag incorrect redshift measurements, or uncertain ones. We provide an in-depth cross-comparison of the various data sets used in this work in Appendix~\ref{Appendix:redshift}. We found  a systematic offset by $\Delta z = -3.6\times10^{-4}$ of the ESO-based data (i.e. FORS and MUSE) compared to GMOS and \cite{Momcheva2015} spectra. While the origin of this offset remains unknown, we have decided to correct the ESO-based measurement for this analysis. In addition, Appendix~\ref{Appendix:redshift} lists the four objects for which we suspect a necessary revision of the published redshift. 

\begin{figure}
	\includegraphics[scale=0.54]{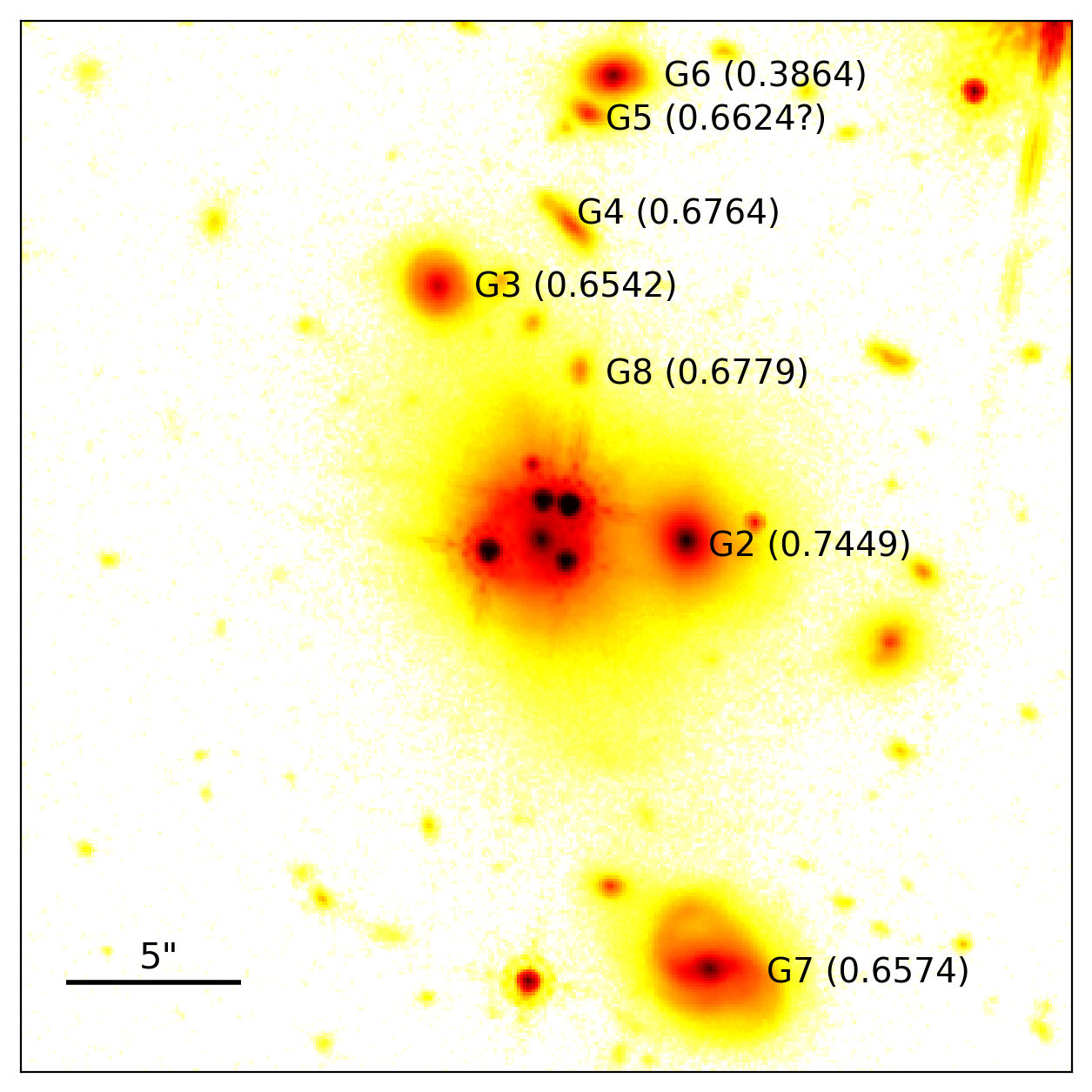}
	\caption{Central 30\arcsec\,$\times$30\arcsec\, region centred on \WFItwenty\,(matching the central black box in Fig.~\ref{fig:field}), with galaxy naming scheme G2-G6 following Vuissoz et al. (\citeyear{Vuissoz2008}) and G7-G8 are our own designation. North is up, East is left. Redshifts (see Sect.~\ref{subsec:specz}) are indicated in parentheses. Insecure redshifts are followed by a question mark.  }
	\label{fig:zoom}
\end{figure}

\begin{table*}
	\caption{Excerpt of the spectroscopic redshift catalog. Columns \#1 to \#6 are objects name (=filename of the 1D spectrum), IDs, positions (RA-DEC, J2000), redshifts $z$ and their uncertainty $\sigma_z$. The last two columns display a quality flag and the object type. The full table is available in electronic form. \label{tab:catalog}}
	\begin{tabular}{lccccccl}
		\toprule
		Name$^{1}$ &   ID &       RA &      DEC &         $z$ &     $\sigma_z$ &  zQF$^{2}$ &       Type$^{3}$ \\
		\midrule
FORS\_20130531\_obj1035 &  1035 &  308.404362 & $-$47.392193 &  0.5381 &  0.0002 &    0 &    Starburst \\
FORS\_20130531\_obj967 &   967 &  308.414562 & $-$47.383083 &  0.1807 &  0.0002 &    0 &    ETG-Sx \\
FORS\_20130531\_obj570 &   570 &  308.431962 & $-$47.385453 &  0.6174 &  0.0002 &    0 &    Starburst \\
FORS\_20130531\_obj445 &   445 &  308.470162 & $-$47.401903 &  0.4434 &  0.0002 &    0 &    Starburst \\
FORS\_20130531\_obj846 &   846 &  308.424862 & $-$47.369983 &  0.3870 &  0.0002 &    0 &    ETG-Sx \\
		\bottomrule
	\end{tabular}
	\begin{flushleft}
		{\small {\bf {Notes:}} $(1)$ Format: Instrument\_date\_objID, where instrument is \emph{FORS}, \emph{Gemini} or \emph{MUSE} if the redshift is derived from our survey, and Momcheva if the redshift comes from \cite{Momcheva2015}. The ``date'' in format yyyymmdd is the date of observation, or \emph{201508} for objects from  \cite{Momcheva2015}. This is also the name of the 1D extracted spectrum.  \\
			$(2)$ The quality flags zQF=0/1/2 if the redshift is extracted from this program and 3,4,5,6 refer to objects from \cite{Momcheva2015}. zQF=0 for secure redshift; zQF=1 for tentative redshift; zQF=2 for unreliable/unknown redshift; zQF=3 for data obtained with LDSS-3; zQF=4 for data obtained with IMACS; zQF=5 for data obtained with Hectospec; zQF=6 for NED objects. \\
			$(3)$ Type=ETG-Sx if CaK-H and/or G-band are detected; Type=Starburst if clear emission lines are observed, Type=M-dwarf for a M-dwarf star; Type=Star for other stellar-types; Type=Unknown if no identification could be done or if the spectrum is from an external catalog.}
	\end{flushleft}
\end{table*}

%

\subsection{Completeness of the spectroscopic redshifts}
\label{subsec:completeness}

\begin{figure}
	\includegraphics[width=1.0\linewidth]{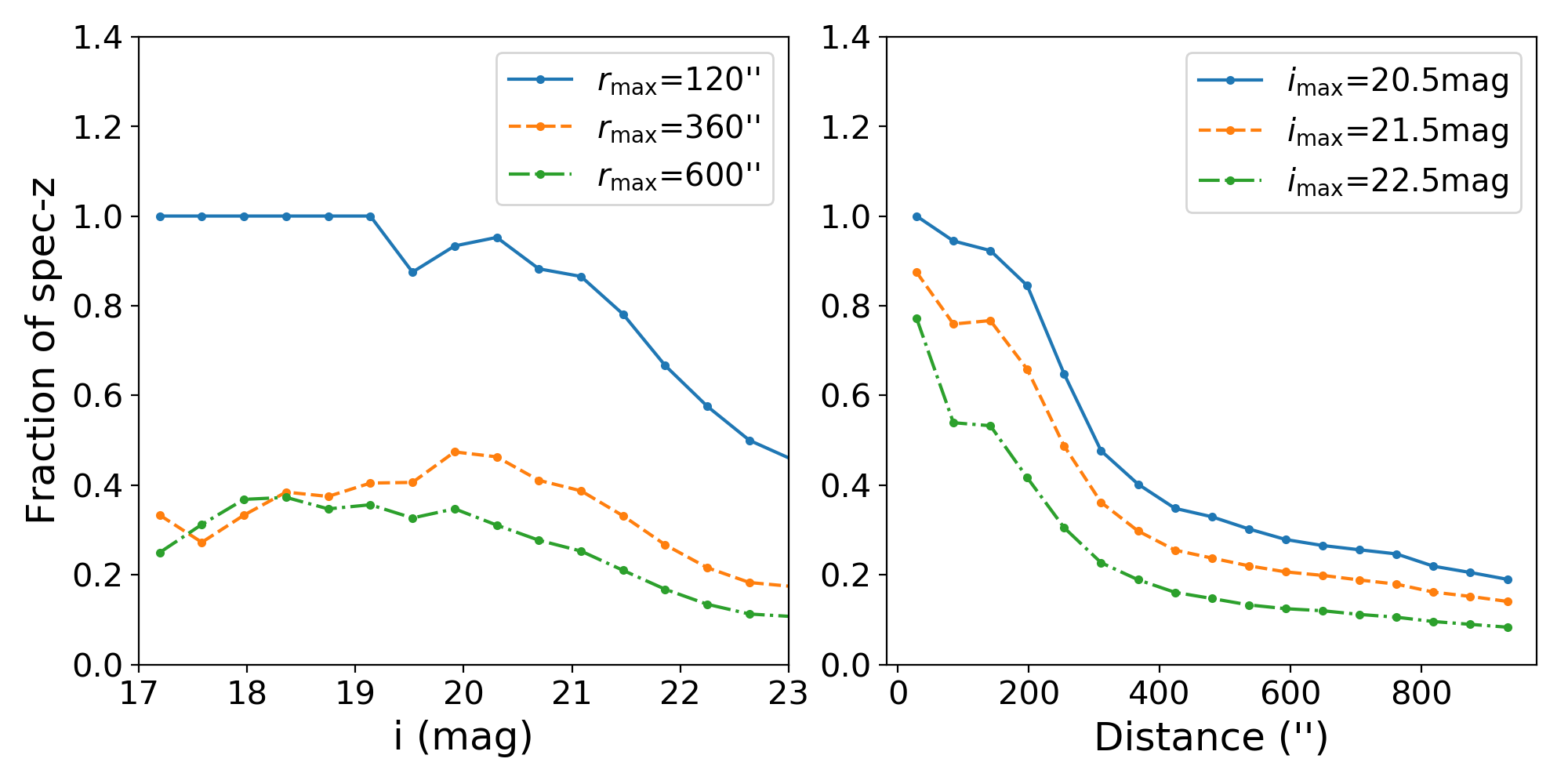}
	\caption{{\bf Left:} Fraction of spectroscopic redshifts of galaxies used in this work (only robust measurements are included) as a function of the maximum $i$-band magnitude of the sample, for three different radii $r_{\rm{max}}$ of 2\arcmin\,(solid-blue), 6\arcmin\,(dashed-orange), 10\arcmin\,(dashed-dotted-green). The low apparent completeness for the brightest objects (for $r_{\rm {max} }> 2$\arcmin ) is caused by several stars mistakenly classified as galaxies in the photometric catalog. {\bf Right:} Fraction of spectroscopic redshifts as a function the maximum distance to the lens for three different limiting magnitude ($i_{\rm max} = 20.5$\,mag (solid-blue); $i_{\rm max} = 21.5$\,mag (dashed-orange), $i_{\rm max} = 22.5$\,mag (dotted-dashed-green).  }
	
	\label{fig:completeness}
\end{figure}

We evaluate the spectroscopic redshift completeness as a function of various criteria by comparing our spectroscopic and photometric catalogs. Figure~\ref{fig:completeness} displays the completeness of our spectroscopic catalog as a function of the limiting magnitude of the sample (fixing the separation to the lens) and of the separation from the lensing galaxy (fixing the limiting magnitude). We see that our completeness is larger than 60\% at small radius, down to $i\sim 22.5$\,mag. This is similar to the completeness reached for the analysis of \HEofor~\citep{Sluse2017}. However, owing to the MUSE data, we have a higher success in the spectroscopic identification of faint sources located at low projected angular separation from the lens. This is particularly important as those galaxies are most likely to produce high-order perturbations at the lens image position.  

\begin{figure}
	\includegraphics[width=1.0\linewidth]{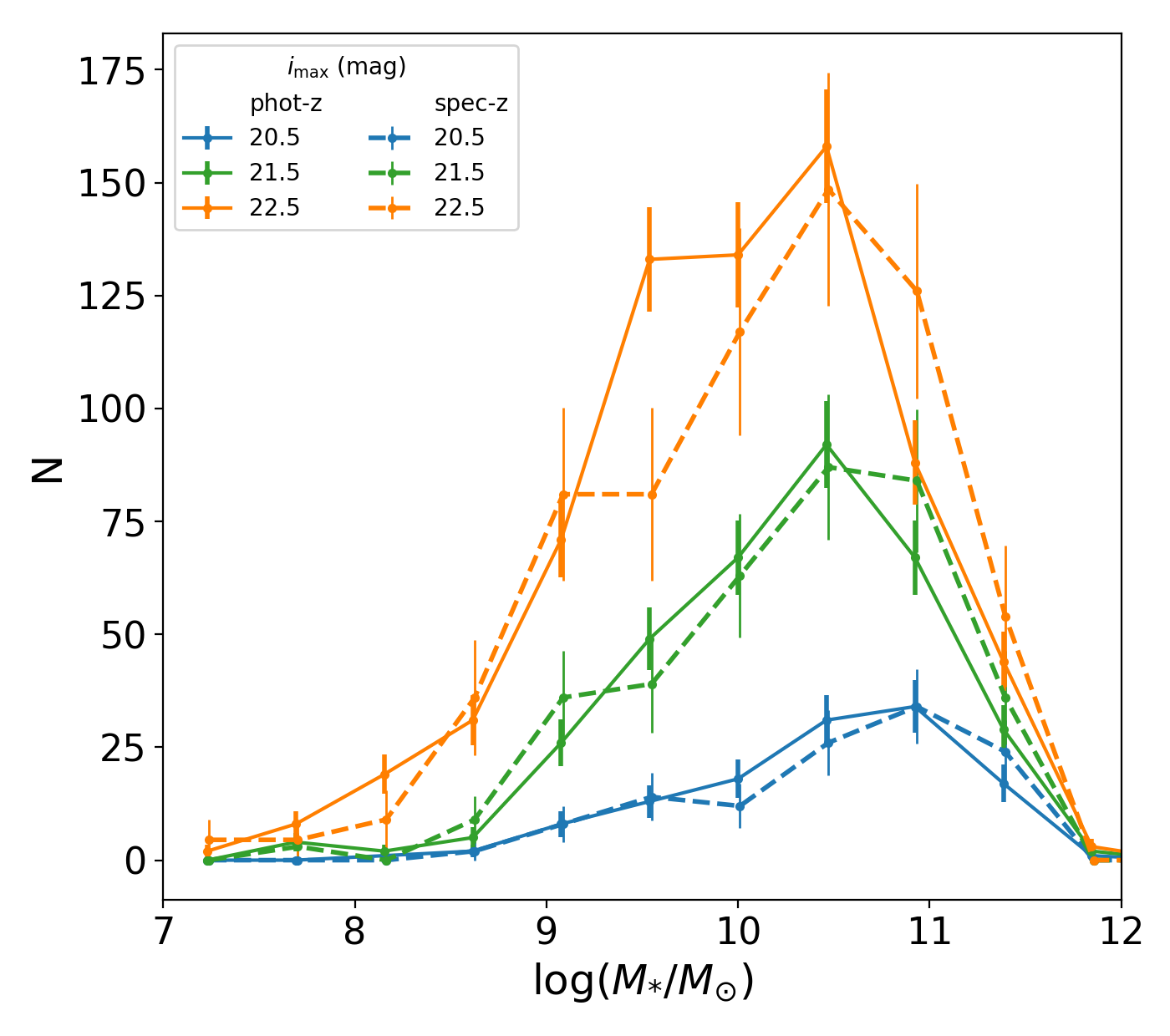}
\caption{Characteristics of the spectroscopic sample for galaxies located less than 6\arcmin\,from \WFItwenty.
	Number of galaxies as a function of the stellar mass for the photometric (solid) and spectroscopic (dashed) samples for three different cuts in magnitudes $i_{\rm max} = $ (21.5, 22.5, 23.5) mag (resp. blue, green, orange). A bin width $\delta(\log(M/M_{\odot})) = 0.5$ has been considered. To ease legibility, for each magnitude cut, the peak of the distribution of the spectroscopically confirmed galaxies has been normalized by a factor $n=(2.0, 3.0, 4.5)$ to match the corresponding peak (i.e. $i_{\rm max} = $ (21.5, 22.5, 23.5) mag) of the photometric sample. In addition, line-plot instead of bar-plot has been used for clarity. } 
\label{fig:Mass}	
\end{figure}

Figure~\ref{fig:Mass} compares the distributions of galaxies (located in projection less than 6\arcmin\, from the lens) in the spectroscopic and photometric samples, as a function of their median stellar mass (as derived in Sect.~\ref{sec:photozmstar}). We see that the two distributions agree well, with a slight over-representation of the most massive galaxies ($M \geq 10^{11}\,M_{\odot}$) in the spectroscopic sample. This is expected as we have a flux limited sample, and more easily measure redshifts of the brightest galaxies. This means that our completeness is the highest for the most massive galaxies, which are also the most likely to perturb the lens gravitational potential. There are no galaxies with $M \geq 10^{11}\,M_{\odot}$ within 1\arcmin\, radius of the lens that are missing spectroscopic redshifts, and only 3 of 12 galaxies if we look up to 2\arcmin\, separation from the lens. Since those 3 galaxies are all located at more than 100\arcsec\, from the lens, this ensures that no massive perturber lacks a spectroscopic redshift.

\section{Galaxy group identification}
\label{sec:groups}

The strategy used to identify groups towards \WFItwenty~is the same as the one developed by \cite{Sluse2017}, building on earlier algorithms implemented in e.g. \cite{ Wilson2016}. We summarize in Sect.~\ref{subsec:group_method} the key aspects of the procedure and refer to \cite{Sluse2017} for more details. Results of the group-finding algorithm are presented in Table~\ref{tab:groups} and Sect. ~\ref{subsec:group_results}. Discussion and comparison with a previous search for groups towards \WFItwenty~are presented in Sect.~\ref{subsec:group_discussion}.

\subsection{Method}
\label{subsec:group_method}
After an iterative filtering in redshift space to identify potential group members, an iterative procedure accounting for the 3D-position of each galaxy is used to refine group membership and estimate the group velocity-dispersion. In practice, we first select a region of angular radius $\theta_{\rm max}$ centred on the lens, bin the redshift catalog in uniform bins of 1000 \kms (i.e. expected maximal velocity-dispersion of a line-of-sight structure), and identify a redshift peak as a bin of more than $N$ elements. The operation is repeated after shifting the bins by half the bin width (i.e. 500 \kms). Then, a first preselection (Step \#1) of potential group members is performed iteratively for each redshift peak. This is realised by building a core subsample of galaxies that only contains those galaxies separated by less than $\delta v_{\rm max} $ from a redshift peak. At each iteration we add galaxies separated by less than $\delta v_{\rm max}$ from the average redshift of this core group, and update the group redshift and velocity-dispersion using a bi-weight estimator. If the new group redshift is found to be more than 2$\times \delta v_{\rm max}$ from the estimated redshift, we restrict our search to 2 $\delta v_{\rm max}$ around the guessed redshift. Our past experience \citep{Sluse2017} suggests that considering $\delta v_{\rm max} = 1500$\kms\, allows one not to miss large distant groups whose effect could be important on the cosmological inference. Based on this filtered galaxy catalog, we refine group membership (Step \#2), accounting for the 3D positions of the galaxies, implementing the method proposed by \cite{Wilman2005}. The algorithm selects galaxies located within $n$ times ($n=2$) a presumed velocity-dispersion ($\sigma_{\rm{obs}} = 500$\,\kms at the first iteration) along the line of sight (i.e. redshift space), and with a maximum aspect ratio between the transverse and radial extension of $b=3.5$. The maximum extension of the group is deduced from the maximum separation between the group centroid (optionally luminosity weighted) and the candidate galaxy members. The galaxies chosen based on these criteria are used to refine the velocity-dispersion $\sigma_{\rm{obs}}$. A gapper algorithm \citep{Beers1990} is used to evaluate $\sigma_{\rm{obs}}$ when there are fewer than 10 galaxies, the dispersion between the velocity measurements when this number drops below 5, and a bi-weight estimator otherwise. This estimate of $\sigma_{\rm{obs}}$ serves as an updated proxy of the velocity-dispersion used to run a new iteration. The algorithm stops when a stable number of group members is found. It also happens that the number of members falls to zero, especially when galaxies are too spread in 3D space (hence not forming a gravitationally linked group). In this situation, no group is associated to the identified redshift peak. 

\subsection{Results}
\label{subsec:group_results}
We have carried out our group search around redshift peaks of $N\geq5$  when $\theta_{\rm max} = 360\arcsec$, and $N\geq10$ when $\theta_{\rm max} = 900\arcsec$. The use of two different $\theta_{\rm max}$ allows us to avoid missing the identification of a small compact group located close to the lens, if another structure at slightly different redshift (i.e. a few thousands km/s) is present at larger radius. The difference of cut-off to identify a peak occurs because, at large distance from the lens, we are only interested in identifying the largest groups that could affect directly the lens-modelling. Table~\ref{tab:groups} lists the properties of the group candidates. A visual inspection of the automatically detected groups revealed that the algorithm tends to identify multimodal distributions in redshift space as a single large structure, yielding group candidates with characteristics of a galaxy cluster (i.e. $\sigma_{\rm{obs}} \sim $ 1000 \kms). In such situations, following \cite{Munoz2013}, we run our algorithm around each redshift peak but restricting the search to $\delta v_{\rm max}= 500 $\kms\, during Step \#1, which is also the typical width of the observed modes in the redshift distribution. The drawback of this approach is that the small groups identified this way generally remain unchanged after step \#2, even when only very few galaxies fall in projection within 1 angular virial radius from the group centroid. Consequently, we manually flag those groups as spurious when fewer than 2 galaxies fall within one angular virial radius from the group centroid. The group centroid is expected to fall close to the brightest galaxy group \citep{Robotham2011, Shen2014a, Hoshino2015}. Since the use of a luminosity weighting does not improve the match between the group centroid and the brightest galaxy (see Appendix~\ref{sec:groups_lweight}), we ignore the latter in the remaining parts of our analysis.
  
\subsection{Discussion}
\label{subsec:group_discussion}

Wilson et al. (\citeyear[][hereafter WIL16]{Wilson2016}) report the semi-automatic search for groups using a methodology very similar to the one used here. Since our catalog includes the catalog used by WIL16, we may expect to recover their group detection, and/or understand whether some detections were possibly spurious. WIL16 report the automatic detection of 5 groups towards \WFItwenty, two of them (at $\bar{z}_{\rm{group}} = 0.1740$ and $\bar{z}_{\rm{group}} = 0.2629$) being flagged as uncertain as they are located close to the edge of their FOV. We identified 2 groups at these redshifts when $\theta_{\rm max}=900''$, but we removed them from the final list because they contain fewer than 10 members. The three other groups reported by WIL16 are found at a redshift compatible with our groups \texttt{a3}, \texttt{a5} and \texttt{a8}, but the number of group members is larger by typically 30\% in our analysis. The properties of \texttt{a3} and \texttt{a5} agree within error bars with our detection, but not \texttt{a8}. In fact, WIL16 report a group of 5 galaxies at $\bar{z}_{\rm{group}} = 0.6838$, namely located at $\sim$1200 \kms~from \texttt{a8} and -1500 \kms~from \texttt{a9}. Our algorithm also originally identified a group candidate of  20 galaxies with $\sigma = 1030$\,\kms~centred at the same redshift as Wilson's group (i.e. $\bar{z}_{\rm{group}} = 0.6840$), but that group candidate has been broken down into \texttt{a8} and \texttt{a9} as the redshift distribution is bimodal, which is not expected in the case of a single group. 
  
In addition to automatic detections, WIL16 report 3 visually identified groups at $\bar{z}_{\rm{group}} =0.3288$, $\bar{z}_{\rm{group}} =0.3926$ and $\bar{z}_{\rm{group}} =0.5100$, as well as  2 groups of fewer than 5 members at $\bar{z}_{\rm{group}} = 0.2151$ and $\bar{z}_{\rm{group}} = 0.3986$. The groups at $\bar{z}_{\rm{group}} =0.3926$ and $\bar{z}_{\rm{group}} = 0.3986$ found by WIL16 may be part of the large over-density of galaxies observed at $z \sim 0.394$ (i.e. 39 galaxies with $z \in [0.382, 0.406]$, or $\pm$ 3500~\kms from $z \sim 0.394$). The distribution of redshifts in that range is multi-modal, suggesting that it is not caused by a massive galaxy cluster\footnote{The visual inspection of 2$\times$15\,ks archive Chandra ACIS data of \WFItwenty~shows only point-like sources, but no diffuse emission that would be associated with a galaxy cluster}. Instead, we identify up to 3 compact groups (\texttt{a0-1}, \texttt{a0-2}, \texttt{a0-3}), two of them ( \texttt{a0-1} and \texttt{a0-3}) roughly matching the central redshift of the group identified by WIL16. The properties of those groups differ however from those reported by Wilson as our data reveal 18 new galaxies in that redshift range. The other groups reported by WIL16 are found by our algorithm when $\theta_{\rm max} = 900\arcsec$, but have been removed because of our choice to only keep groups of at least 10 members for large $\theta_{\rm max}$. The properties of those groups, while not identical to those of WIL16 due to the higher completeness of our catalog, are compatible with the groups of WIL16. 

There are two group candidates reported in our work that are absent of WIL16, namely \texttt{b5} and  \texttt{a2}. The group \texttt{a2} hosts galaxies identified exclusively based on our new data sets. It is therefore expected that WIL16 report no detection at that redshift. On the contrary, 6 of the 8 galaxies identified in \texttt{b5} were also in WIL16 catalog. As pointed out in Table~\ref{tab:groups}, the velocity histogram of this group is bimodal, such that its reported properties are likely biased. If split in two, the two sub-groups would miss our threshold of 5 galaxy members to be classified as a group. Our finding is therefore compatible with the lack of detection by WIL16. 

\begin{table*}
	\caption{Properties of the groups identified in the FOV of \WFItwenty. The columns are the group redshift, the number of spectroscopically identified galaxies in the group, the group intrinsic velocity-dispersion (rounded to the nearest 10\,\kms) and 1$\sigma$ standard deviation from bootstrap, the group centroid, bootstrap error on the centroid, projected distance of the centroid to the lens, median flexion shift $\log(\Delta_3 x (\rm{arcsec}))$ and 1$\sigma$ standard deviation from bootstrapping (Sect.~\ref{sec:model}). The last column indicate for which field a peak of more than 5 galaxies is detected in redshift space. The properties we display correspond to the FOV marked in bold. }
	\label{tab:groups}
	\centering
	\begin{minipage}{\linewidth}
		\centering
	\begin{tabular*}{0.9\linewidth}{llrcccccc}
		\hline
		ID & $\bar{z}_{\rm group}$ & N & $\sigma_{\rm{int}}$ (err)& RA$_{\rm ctr}$, DEC$_{\rm ctr}$ & err(RA$_{\rm ctr}$, DEC$_{\rm ctr}$) &  $\Delta\theta$ & $\log(\Delta_3 x) \pm $ err & FOV  \\
		& & & \kms & deg &  arcsec &  arcsec   & $\log(\rm{arcsec})$ & arcmin \\
\hline 
b5 & 0.3060$^\star$ & 8 & 530 (110) & 308.61026100, $-$47.43226275 &  145.2, 15.6  & 469.0 &$-$6.59 $\pm$ 0.48 & 900 \\ 
a0-1 & 0.3937$^\star$ & 7 & 140 (30) & 308.40536200, $-$47.41128275 &  17.1, 8.8  & 75.3 & $-$7.25 $\pm$ 0.58&  360  \\ 
a0-2 & 0.3867$^\star$ & 8 & 100 (20) & 308.42486699, $-$47.36855275 &  78.1, 17.2  & 96.9 & $-$7.50 $\pm$ 0.57&  360 \\
a0-3 & 0.3999$^{\dagger, \star}$ & 12 & 380 (70) & 308.34270600, $-$47.33671275 &  87.4, 79.4  & 292.5  & $-$7.01 $\pm$ 0.40& 360 \\ 
a2 & 0.4436 & 6 & 150 (40) & 308.47654650, $-$47.39469775 &  36.3, 21.1  & 124.2 & $-$7.02 $\pm$ 0.54 & 360 \\
a3 & 0.4956 & 13 & 520 (100) & 308.46337200, $-$47.36336725 &  61.8, 58.9  & 147.8 & $-$4.98 $\pm$ 0.81 & {\bf{360}}, 900 \\
a5 & 0.6588 & 22 & 500 (80) & 308.43557011, $-$47.37411275 &  35.6, 18.6  & 80.7 & $-$4.70 $\pm$ 0.45 & {\bf{360}}, 900 \\
a8 & 0.6796 $^\ddagger$ & 11 & 610 (190) & 308.42531059, $-$47.39318538 &  68.6, 24.3  & 8.3 & $-$3.75 $\pm$ 1.21 & 360 \\
a9 & 0.6889$^\star$ & 4 & 190 (90) & 308.41116200, $-$47.41528275 &  24.9, 17.5  & 79.5 & $-$6.42 $\pm$ 2.26 & 360 \\

\hline
\end{tabular*}
\begin{flushleft}
{\small 
	{\bf Note:} ${\dagger}$ Likely spurious. $\ddagger$ Apparently bimodal but unsuccessful breakdown into sub-group(s). $\star$ Results from the breakdown of a larger multi-modal group candidate. \\
}
\end{flushleft}

\end{minipage}

\end{table*}

\begin{figure*}
	\includegraphics[scale=0.33]{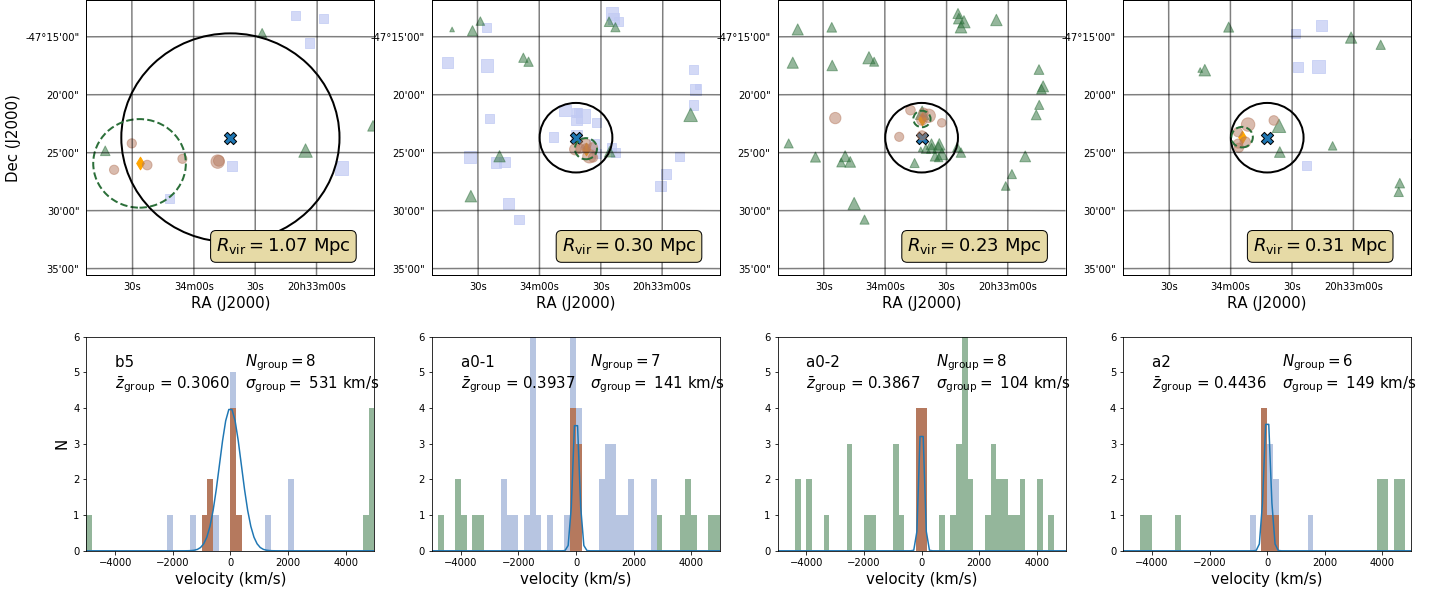}
	
	\caption{Main groups identified in the field of \WFItwenty: For each redshift (column), the distribution of (rest-frame) velocities of the group galaxies identified spectroscopically is shown (bottom panel) together with a Gaussian of width equal to the intrinsic velocity-dispersion of the group. Bins filled in red correspond to galaxies identified as group members, in blue as interlopers in redshift space, and in green as non-group members. The top panel shows the spatial distribution of the galaxies with a redshift consistent with the group redshift, using the same color scheme as for the bottom panel. The positions of the lens (group) centroid is indicated with a cross (orange diamond). The size of the symbol is proportional to the brightness of the galaxy, and color code is the same as for the bottom panel. The solid (dashed) black (green) circles show the field used to identify the peak initial guess for the group redshift (a field of radius $r\sim 1 \times R_{\rm vir}$). The groups with the largest flexion shifts (and hence, potentially the largest impact on the modelling) are the groups $a3$, $a5$ (that includes the lens), and $a8$ (see Sect. ~\ref{subsec:flex_groups}, and continued panels of this figure). }
	\label{fig:maingroups}
\end{figure*}

\begin{figure*}
	\includegraphics[scale=0.33]{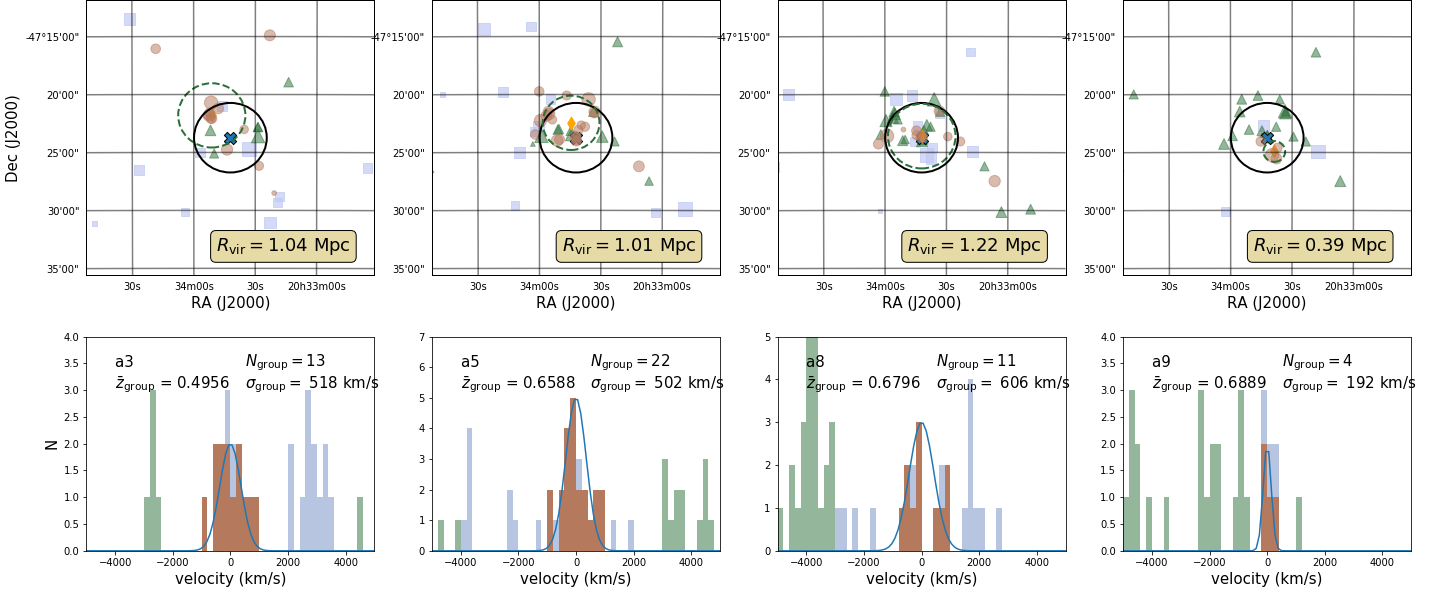}

		\setcounter{figure}{5}
	\caption{continued. }
\end{figure*}

\section{ Contribution of line of sight and environment to the lens structure}
\label{sec:model}

We are interested in identifying the structures (galaxies or galaxy groups) that require explicit modelling in the course of the cosmological inference, but may not be accounted for using a tidal approximation. For that purpose, we need to identify massive galaxies or groups that fall too close in projection to the lens to produce only a uniform perturbation of the main lens gravitational potential over the area covered by the lensed images. As in \citet{Sluse2017}, we use the diagnostic proposed by \cite{McCully2014, McCully2017}. The method consists of comparing the shift of the solutions of the lens equation with and without including the flexion produced by the perturber (a single galaxy or a galaxy group). For a point mass, the magnitude of the shift produced by the flexion term, called ``flexion shift'' $\Delta_3 x$, can be written:
\begin{equation}
\Delta_3 x = f(\beta) \, \times \frac{(\theta_{\rm E} \,\theta_{\rm E,p})^2}{\theta^3}, 
\label{eq:flexion}
\end{equation}

\noindent where $\theta_{\rm E}$ and $\theta_{\rm E, p}$ are the Einstein radii of the main lens and of the perturber, and $\theta$ is the angular separation on the sky between the lens and the perturber. We define $f(\beta) = (1-\beta)^2$ if the perturber is behind the main lens, and $f(\beta) = 1$ if the galaxy is in the foreground. Here, $\beta$ is defined for a galaxy at redshift $z_{\rm p} > z_{\rm d}$ as:
\begin{equation}
\beta = \frac{D_{\rm{dp}} D_{\rm{os}}}{D_{\rm {op}} D_{\rm{ds}}},
\end{equation}

\noindent where the $D_{ij} = D(z_i, z_j)$ corresponds to the angular diameter distance between redshift $z_i$ and $z_j$, and the subscripts o, d, p, s stand for the observer, deflector, perturber, and source. We explain in the next section how $\theta_{\rm E,p}$ is determined. 

As long as the flexion shift of a galaxy is (much) smaller than the observational precision on the position of the lensed images, its perturbation on the gravitational potential of the main lens can be neglected in the lens model. Based on the simulation results of \cite{McCully2017}, we adopt the likely conservative threshold of $\Delta_3 x >  10^{-4}$\, arcseconds, i.e. more than 10 times smaller than the astrometric accuracy of the data used in the cosmological inference analysis. Those authors show that by considering explicitly galaxies or galaxy groups with flexion shift larger than this threshold, we limit the bias on $H_0$ at the percent level in the cosmological analysis. 

\subsection{Individual galaxies}
\label{subsec:flex_galax}

We first calculate the flexion shift for the individual galaxies in the field of \WFItwenty. This requires an estimate of the Einstein radius $\theta_{\rm E, p}$ of these galaxies. This is achieved in a two-step process. First, we infer the line-of-sight central velocity-dispersion $\sigma_{\rm{los}}$ of each galaxy using the scaling relation from \cite{Zahid2016}, and DES-based stellar masses (Sect.~\ref{sec:photozmstar}). This empirical "double power-law" relationship has been derived from a large sample of early-type galaxies at $z < 0.7$ observed with SDSS, covering the stellar mass range $\log(M_{\star} / M_{\odot}) \in [9.5, 11.5]$. Since no significant modification of the relationship has been found by \cite{Zahid2016} when splitting the sample in different redshift bins, we assume no evolution with redshift. In addition, we assume that this relationship is still valid at the low-mass end of our sample, where $M_{\star} < 10^{9.5} M_{\odot}$. In a few cases, when no accurate multi-band photometry was available due to object blending, we fix the stellar mass to $10^{10.17} M_{\odot}$, namely the median stellar mass of the whole sample. We use the relation from \cite{Zahid2016} without regard to the galaxy type. This is a conservative choice as, for the same luminosity, early-types have a larger velocity-dispersion than spirals. Therefore, we may only overestimate the flexion from individual galaxies. 

Second, we adopt a Singular Isothermal Sphere to convert the velocity-dispersion of the galaxy into its Einstein radius $\theta_{\rm{E,p}}$: 

\begin{equation}
\theta_{\rm E, p} = 4\pi \left(\frac{\sigma_{\rm los}}{c}\right)^2 \, \frac{D_{\rm{ps}}}{D_{\rm{os}}}, 
\label{eq:SIS}
\end{equation}

\noindent where $D_{\rm{ps}}$ ($D_{\rm{os}}$) is the angular diameter distance between the perturber $p$ (resp. the observer $o$) and the source $s$. All along the procedure, we use the spectroscopic redshift if available, and the photometric redshift otherwise to calculate distances, together with the stellar mass computed in Section \ref{sec:photozmstar} at this corresponding redshift. Table~\ref{tab:flexion_galax} lists the 10 galaxies with the largest flexion shifts. Only four of them have a flexion shift $\Delta_3 x > 10^{-4}~$arcsec, namely the galaxies labeled $G2, G3, G7$ and $G8$ on Fig.~\ref{fig:zoom}. Among those galaxies, $G8$ does not have reliable multi-band photometry, and therefore a stellar mass of $\log(M / M_{\odot}) = 10.17$ has been assumed. This arbitrary choice may yield a substantial overestimate of the flexion shift. Indeed, this galaxy shows spectroscopic characteristics of a spiral galaxy, and is clearly fainter than $G4$, another spiral located at about the same redshift than $G8$, but with a photometric stellar mass of only $\log(M / M_{\odot}) \in  [8.96, 9.48]$. Assessing a stellar mass in that range for $G8$ yields flexion shifts $\Delta_3 x \in $ [$3.43\,10^{-5}$, $2.55\,10^{-5}$] arcsec, well below the threshold above which that galaxy would have a substantial impact on the modelling. 

The uncertainty on the flexion shift of each galaxy is derived by quadratically adding the uncertainty originating from the conversion of stellar mass into $\sigma_{\rm{los}}$, with the uncertainty on the stellar mass itself (which is strongly correlated with the photometric redshift, such that we can effectively neglect the redshift uncertainty). More precisely, we calculate the flexion using the 16 and 84 percentile values uncertainty on the velocity-dispersion from Fig.~9 of \cite{Zahid2016} to derive the $1~\sigma$ uncertainty originating from the velocity-dispersion; and we calculate the uncertainty originating from the stellar mass by calculating the flexion using the 16 and 84 percentile values of the stellar mass. Those two contributions to the error budget yield a typical $1~\sigma$ uncertainty of 0.5 dex on $\log(\Delta_{3}x / {\rm arcsec})$. 

\begin{table*}
	\caption{Main characteristics of the 10 galaxies with the largest flexion shift. The first 6 columns display the galaxy ID (and label used in Fig.~\ref{fig:zoom} if displayed), coordinates (RA, DEC in degrees; ICRS), redshift $z$, $i$-band magnitude, and distance to the lensing galaxy (in arcsec). The next three columns provide the logarithm of the flexion shift $\log(\Delta_{3}x / 1 \rm{arcsec})$ for three different percentiles of the posterior distribution, i.e. 16, 50 and 84 percent (see Sect.~\ref{subsec:flex_galax} for details). Values of the flexion shifts $\Delta_{3}x > 10^{-4}$\,arcsec are displayed with bold font to ease the identification of the most prominent perturbers.}
	\label{tab:flexion_galax}
\begin{tabular}{ccccccccc}
\hline
ID & RA & DEC & $z$ & MAG & dist & $\log(\Delta_{3}^{16}x)$ & $\log(\Delta_{3}^{50}x)$ & $\log(\Delta_{3}^{84}x)$ \\
\hline
501 (G2) & 308.424014 & $-$47.395599 & 0.7449 & 20.02 & 3.8 & {\bf $-$2.65} & {\bf $-$1.99} & {\bf $-$1.60} \\
1100 (G8)$^\dagger$ & 308.425195 & $-$47.394358 & 0.6779 & -         & 4.1 & $-$4.07 & {\bf $-$3.51} & {\bf $-$3.09} \\
482 (G7) & 308.423727 & $-$47.398862 & 0.6574 & 20.37 & 13.0 & $-$4.58 & {\bf $-$3.91} & {\bf $-$3.54} \\
581 (G3) & 308.426804 & $-$47.393648 & 0.6542 & 21.19 & 7.2 & $-$4.98 & $-$4.16 & {\bf $-$3.71} \\
1045 (G6) & 308.424872 & $-$47.392103 & 0.3864 & 21.28 & 12.3 & $-$5.08 & $-$4.41 & $-$4.01 \\
967 & 308.414562 & $-$47.383083 & 0.1807 & 17.96 & 52.1 & $-$5.16 & $-$4.60 & $-$4.32 \\
468 & 308.424933 & $-$47.400595 & 0.6588 & 21.01 & 18.5 & $-$5.47 & $-$4.72 & $-$4.29 \\
574 & 308.435791 & $-$47.391868 & 0.6845 & 20.88 & 28.1 & $-$5.55 & $-$5.00 & $-$4.63 \\
567 & 308.420762 & $-$47.384463 & 0.6574 & 20.71 & 41.3 & $-$5.77 & $-$5.28 & $-$4.93 \\
344 & 308.429001 & $-$47.412962 & 0.6170 & 20.05 & 63.5 & $-$5.68 & $-$5.29 & $-$5.02 \\

\hline
\end{tabular}
\begin{flushleft}
	{\small 
		{\bf Note:} ${\dagger}$ Flexion shift likely over-estimated due to lack of photometric measurement. \\
	}
\end{flushleft}
\end{table*}

\subsection{Groups} 
\label{subsec:flex_groups}

Because galaxies of a group reside in a common dark matter halo, it is important to assess whether the groups identified in Sect.~\ref{sec:groups} need to be explicitly accounted for in the model by attaching a specific mass distribution to their observed centroid. Similarly to the methodology used for the galaxies, we adopt the flexion-shift $\Delta_3 x$ (Eq.~\ref{eq:flexion}) as an indicator of the impact of each group on the model. By describing the group as a singular isothermal sphere, we can calculate the group's Einstein radius (Eq.~\ref{eq:SIS}) based on its velocity-dispersion, and hence $\Delta_3 x$ for each group (Table~\ref{tab:groups} and Table~\ref{tab:groups_lweight}). 

In order to account for the uncertainty on the group centroid and velocity-dispersion, we have repeated the flexion shift estimate on 1000 bootstrap samples of these quantities. More specifically, we resample with replacement the identified group members (i.e. their position and redshifts) and recalculate the group properties using the resampled members. We calculate the flexion shift for each bootstrap group and estimate the 16 and 84 percentiles based on the bootstrapped distribution. We have conservatively considered that groups for which $\Delta_3 x > 10^{-4}$\,arcsec for more than 5\% of the bootstrap samples need to be scrutinized. We discuss below the properties of the these groups:  
\begin{itemize}
	\item \texttt{a3} at $\bar{z}_{\rm group} = 0.4956$. The group centroid falls in the vicinity of a subset of 5 galaxies located within less than 20\arcsec~in projection from the lens. One of those galaxies is the second-brightest galaxy of the group candidates, the brightest one being located in the outskirts of the group (in projection). 

	\item \texttt{a5} at $\bar{z}_{\rm group} = 0.6588$: This group hosts the lensing galaxy. The group properties have only a very weak dependence on the weighting scheme used to estimate the group centroid. The latter is distant by about 80\arcsec~ from several group members, none being the brightest group galaxy.  
	
	\item \texttt{a8} at $\bar{z}_{\rm group} = 0.6796$: the distribution in velocity space for this group is very clumpy. This strongly suggests that this group candidate is a spurious detection, as reported in Tables~\ref{tab:groups} \&  \ref{tab:groups_lweight}. For that reason, we have decided to discard this group in the lens models used for cosmological inference (\rusup).
\end{itemize}

In addition to those groups, we have also estimated the flexion shift caused by the group of 5 galaxies at  $\bar{z}_{\rm{group}} = 0.6840$ identified by Wilson et al. (\citeyear[][see Sect.~\ref{sec:groups}]{Wilson2016}). We find $\Delta_{3}x = 1.8\times10^{-7}$ arcsec, supporting the small impact of this group candidate on the modelling . 
 
\section{velocity-dispersions of individual galaxies}
\label{sec:vdisp}

The velocity-dispersion provides a means of measurement of a galaxy mass. Including this information in the lens-modelling allows us to improve the accuracy of the lens models \citep{Treu2002, Treu2002b, Koopmans2004, Shajib2017}. In addition to the lensing galaxy $G$, the three galaxies with the largest flexion shift (i.e. $G2$, $G3$, $G7$; see Fig.~\ref{fig:zoom} and Tab.~\ref{tab:flexion_galax}) are bright enough to enable a velocity-dispersion measurement with MUSE data. 

For that purpose, we use a code that reproduces an observed galaxy spectrum by performing a Bayesian exploration of the stellar population of the galaxy \citep{Auger2009}. More precisely, we model the observed spectrum as a linear combination of stellar spectra multiplied by a sum of orthogonal polynomials (to account for imperfect sky subtraction and uncertainties in the absolute calibration of the spectrum), convolved with a Gaussian kernel to mimic the line-spread-function of the instrument. Contrary to \cite{Auger2009}, which uses synthetic stellar spectra, we use an ensemble of real stellar spectra of various types and temperatures (i.e. A0, F2, G0, G5, G8, K1, K2 -III stellar types) from the Indo-US spectral library \citep{Valdes2004}. Those spectra  cover the rest-frame range [3465-9469]\,\AA, with a constant spectral resolution of 1\,\AA\, that correspond to $\sigma_{\rm{template}} \sim 28$\,\kms over the wavelength range considered. The instrumental line-spread-function, derived based on a third order polynomial fit of the spectral resolution with wavelength, is characterized by a median FWHM $ \sim 55$\,km/s over the wavelength range considered. The parameters of the models are the coefficients of the polynomial function that accounts for uncertainties on the flux calibration (nuisance parameters), the coefficients of the linear combination of stellar spectra, the velocity offset compared to the guess redshift, and the velocity dispersion. The priors are uniform for all these parameters, with a range limited to [-350, +350]\,\kms\, for the velocity, and [5, 350]\,\kms\, for the velocity dispersion. This methodology, already successfully applied in \cite{Auger2009, Suyu2010, Sonnenfeld2012}, is optimized for measuring velocity-dispersion of spectra with SNR $\geq$ 10 per pixel. 

A brute force extraction of the spectra of $G$ and $G2$ yields significant contamination of the galaxy spectra by the lensed quasar images, precluding a robust velocity-dispersion measurement. It is therefore necessary to remove the quasar flux and local sky residual prior to the velocity-dispersion measurement. For that purpose, we first model each slice of the data cube containing the lens system and $G2$ using a 2D model of the light distribution. This model is constituted of the sum of several components: (1) a de Vaucouleurs light profile, convolved with a PSF model; (2) 4 point-like sources whose relative positions are held fixed to the {\emph{HST}} positions; and (3) a spatially uniform sky background. The code, already successfully used to model ESO-SINFONI IFS data of gravitationally lensed quasars \citep{Braibant2014, Sluse2015}, uses the MPFIT implementation of the Levenberg-Marquardt least-square optimization algorithm \citep{Markwardt2009}. The lowest residuals are obtained using a symmetric Moffat profile for the PSF. The extraction of the galaxy spectrum is performed within a fixed aperture on the data cube after subtraction of the lensed images and sky models. The associated error is calculated by summing the variance of each spaxel in the aperture. 

The velocity-dispersion measurement-steps take as input the spectrum of the galaxy and its associated variance. It is necessary to mask regions of the spectra affected by telluric sky absorption and/or residual sky background not perfectly removed by the reduction procedure, as those features may be mistakenly attributed to stellar features (in a complex way that depends on the object's redshift). For that purpose, we have performed the measurements using different masking schemes (see Appendix~\ref{Appendix:skymask} for details). Multiple aperture radii have been tested for the extraction, and we choose 4 pixels radius (i.e. a square of 9 pixels = 1.8\arcsec\, side-length) as the best compromise between an aperture too small compared to the seeing, and an aperture too large such that the uncertainty on the estimate of the sky subtraction and PSF modelling of nearby targets (quasar images and/or nearby galaxies) contribute to a large fraction of the integrated galaxy flux and introduce a large systematic error on its flux. The final velocity-dispersion measurement (Table \ref{tab:vdisp}) results from the marginalization of the probability distribution function obtained for the different masking schemes, and three choices of polynomial order (i.e. order 3, 4, 5). The confidence interval is defined as the region centered on the median and including 68.4\% of the probability distribution. 

The seeing has been estimated by fitting a Gaussian profile independently for each wavelength slice, on 3 field stars. In that process, we have ignored spectral slices masked out for measuring the velocity-dispersion. In addition, the seeing has also been estimated on the quasar images when modelling the lens-system luminosity profile. We observed an apparent bias in the FWHM measurement caused by sky residuals (FWHM agree between the different stars better for SV data than for P97 data, and the agreement is better at redder wavelengths). Therefore, we use the FWHM of the brighest star as our proxy of the PSF width. The latter agrees with the FWHM derived from the quasar but is systematically larger by 10\%. Because of the more complex measurement of the quasar FWHM, we decided to choose the star-based FWHM as our proxy of the seeing. We measured a median and scatter (along the wavelength direction) of the seeing: FWHM $= 1.06 \pm 0.1 $ for the SV data and FWHM $=  0.98 \pm 0.25$ for the P97 data.

\begin{table}
	\caption{Median velocity-dispersion and 68.4\% CI of 4 galaxies in the FOV of \WFItwenty~(See Fig.~\ref{fig:field}). Measurements are performed within a square aperture of $3 \times 3$ pixels (1.8\arcsec) centred on the galaxy. The last column indicates which MUSE data set has been used for the velocity-dispersion measurement. }
	\label{tab:vdisp}
	\centering
	\begin{tabularx}{0.98\linewidth}{lllll}
		\toprule
		Name & RA,DEC & $\sigma$  & CI on $\sigma$ & Note \\
		& (deg) & (\kms) & (\kms) & \\
		\midrule
		G     & (308.42558, $-$47.39547) &250     &229, 267  & SV      \\
		G2   & (308.42402, $-$47.39560) &232     & 222, 243 & P97    \\
		& &218     & 213, 222 & SV     \\
		& &223     & 215, 237 & P97+SV \\
		G3   &  (308.42680, $-$47.39365) &79      & 60, 102  &  SV      \\
		G7   &  (308.42373, $-$47.39886)  &166     & 160, 173 & SV       \\
		\bottomrule
	\end{tabularx}
	
\end{table}

\begin{figure}
	\includegraphics[scale=0.33]{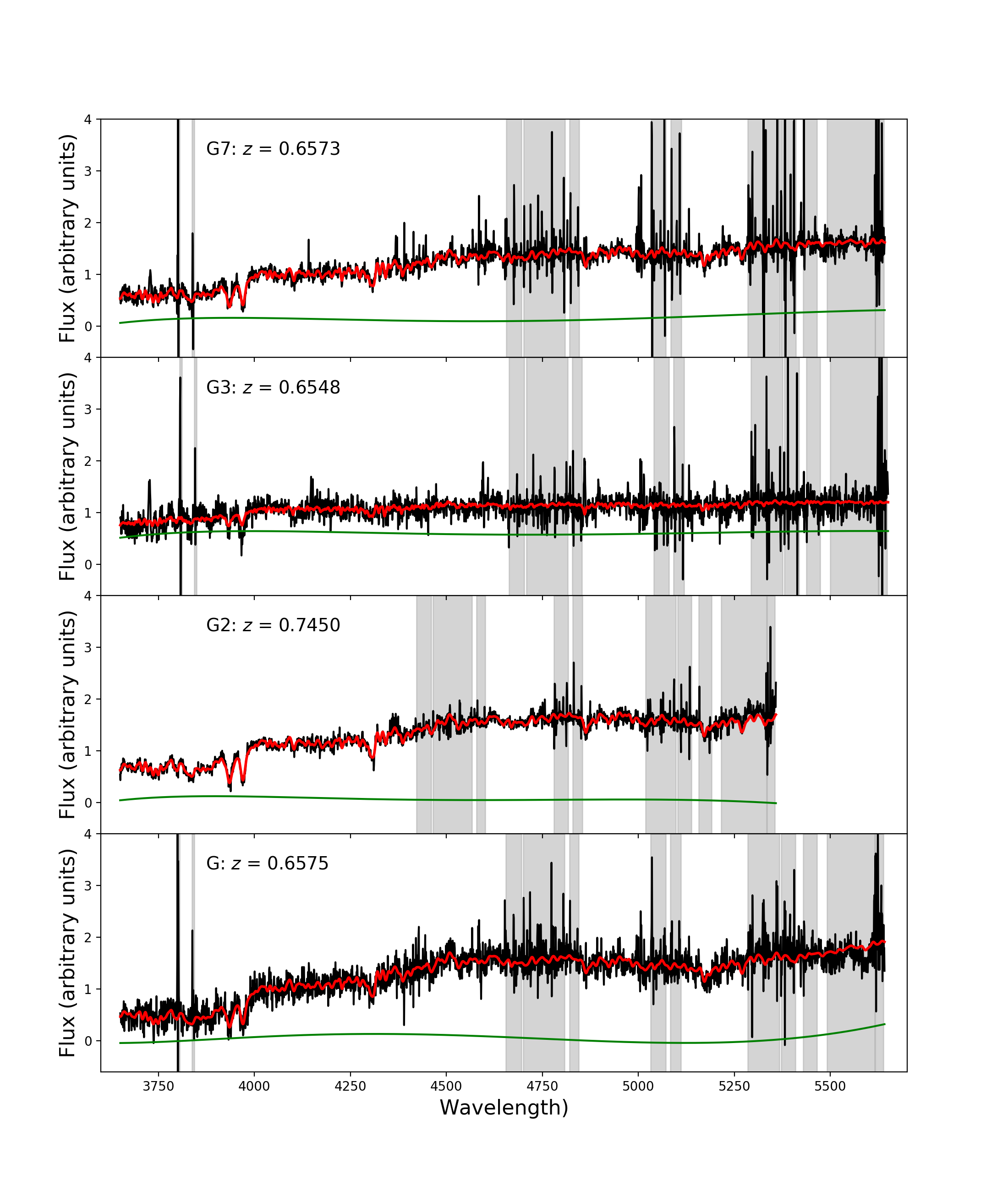}
	\caption{ Rest-frame spectra of the galaxies $G$-$G2$-$G3$-$G7$ (see Fig.~\ref{fig:zoom} for identification) over-plotted with the velocity-convolved synthetic stellar population spectrum (red) used to measure the velocity-dispersion. The grey areas display the regions ignored in the velocity-dispersion fitting process due to the presence of known sky absorption or large variability. The green curve is a multiplicative polynomial of order 4 used to correct mismatch between the observed spectrum and the synthetic one. The redshift, measured simultaneously with the velocity-dispersion measurement, includes a systematic correction by $\Delta z \sim -3.6\times10^{-4}$ (See Appendix~\ref{Appendix:redshift}). Those redshifts agree statistically with those derived in our redshift catalogue (Sect.~\ref{subsec:specz}). }
	\label{fig:vdisp}
\end{figure}


\section{Summary}
\label{sec:conclude}

In the framework of inferring $H_0$ from the time-delay lens system \WFItwenty\,(\bonvinp, \rusup, \wongp), we have performed a detailed characterization of the environmental properties of this system, with the following immediate objectives: (1) identify individual galaxies and galaxy groups susceptible to produce high-order perturbation to the lens potential and therefore requiring to being explicitly included in the lens models and estimate their masses; (2) derive redshift proxies (i.e. spec-zs or photo-zs) for all the galaxies in the field of view to enable a statistical estimate of the convergence associated to the galaxies along the line of sight; (3) measure the velocity dispersion of the main lensing galaxy for self-consistency of the mass modelling.

To reach these goals we have measured photo-zs, spec-zs and inferred stellar masses for most of the galaxies up to $\approx$4\arcmin\, from the lens, down to $i \approx 23.0$\,mag. We have used deep multicolour imaging as well as multi-objects and integral-field spectroscopy. In particular, we used $grizY$ imaging from the Dark Energy Survey, proprietary $u$-band obtained with DECam, near-infrared HAWK-I ($JHKs$) and {\emph {HST}} (F160W), mid-infrared IRAC-Spitzer data, multi-object spectroscopy with ESO-FORS2 and Gemini-GMOS instruments. We have complemented those data with the spectroscopic catalog from \cite{Momcheva2015} who spectroscopically measured the redshift of galaxies distant by up to 15\arcmin\,from \WFItwenty\, down to $i\approx 21.5$. In addition we have also used the exceptional capabilities of the ESO-MUSE integral-field spectrograph to derive spectroscopic redshift of the object closer in projection from the lens (with a projected distance as large as 30\arcsec  from \WFItwenty), but also to obtain velocity-dispersions of the brightest galaxies susceptible to produce high-order perturbation of the lens potential.  With 64 galaxies having a confirmed redshift within a radius of $2$\arcmin\,from the lens, we double the number of systems with a measured spectroscopic redshift in the direct vicinity of \WFItwenty.

Our main results are the following:
\begin{enumerate}
	\item We have gathered a catalog of 366 galaxies with confirmed spectroscopic redshifts in the FOV of \WFItwenty. In addition, we have tentative redshift measurements for 24 galaxies, and 79 objects for which no redshift could be measured. We also spectroscopically  identify 110 stars in the FOV. 
	
	\item We used the same methodology as \cite{Sluse2017} to identify groups of more than 5 (10) galaxies located within 6 (15) arcmin from the lensing galaxy. This selection does not aim at identifying all the groups along the line of sight, but those that are more susceptible affecting cosmological inference with the time-delay method, namely small groups close in projection from the lens, and/or more massive groups/clusters located farther away. Nine group candidates fulfilling those criteria were found, but two of them are likely to be spurious identifications. In particular, a0-3 (see Tab.~\ref{tab:groups}) has fewer than 2 galaxies appearing in projection within one virial radius from its centroid. Another group candidate, a8, shows a bimodal redshift distribution unlikely to be associated with a single group, but our algorithm is unsuccessful in identifying this over-density as 2 separated groups, or one group + isolated galaxies. 
	
	\item We confirm earlier findings  that the main lensing galaxy is part of a large group at $\bar{z}_{\rm{group}} = 0.6588$ \citep{Morgan2004, Wilson2016}, for which we derive $\sigma_{\rm los} = 500 \pm 80$\kms. The number of spectroscopically confirmed members has increased by 30\% owing to this work, and is now reaching 22 galaxies. The lensing galaxy is the seventh brightest galaxy of the group and is therefore suspected not to lie at the centre of its host halo. 
	 
	\item Following \cite{McCully2017}, we have calculated the flexion shift $\Delta_{3x}$ to identify the galaxies/galaxy groups along the line-of-sight most susceptible to produce high-order perturbation of the lensing potential. Two groups may require to be included explicitly in the lens models: The group a3 at $\bar{z}_{\rm{group}} = 0.4956$, for which we identified 13 group members, and the group a5 which hosts the lensing galaxy. In addition, three galaxies ($G2$, $G7$, $G3$, see Fig.~\ref{fig:zoom}) are susceptible to produce a non-negligible high-order perturbation of the main lens gravitational potential. Owing to our MUSE spectroscopic data, we have been able to measure the velocity-dispersion for these galaxies, which are used as a prior for the lens-modelling of \WFItwenty\, presented in \rusu.  
	
	\item We measure the velocity-dispersion of the lens to be $\sigma_{\rm{los}}= 250^{+15}_{-21}$ \, \kms. 
	
\end{enumerate}

These results are used by \rusu\,to account for the main perturbers explicitly in the mass modelling of \WFItwenty\,, quantify the statistical contribution to the main lens potential of galaxies along the line of sight, and constrain $H_0$ from the time delay measured in that lens system (\bonvinp). \wong\, present the constraints on various cosmological parameters combining the H0LICOW lenses analysed so far. 


\section*{Acknowledgments}

We thank Malte Tewes for his contribution to the \HOLI~project. C.E.R. wishes to thank A. Tomczak for providing the PSF matching code. \HOLI~and COSMOGRAIL are made possible thanks to the continuous work of all observers and technical staff obtaining the monitoring observations, in particular at the Swiss Euler telescope at La Silla Observatory. Euler is supported by the Swiss National Science Foundation. 
This project has received funding from the European Research Council (ERC) under the European Union’s Horizon 2020 research and innovation programme (grant agreement No 787886). 
This work was supported by World Premier International Research Center Initiative (WPI Initiative), MEXT, Japan.
T.T. thanks the Packard Foundation for generous support through a Packard Research Fellowship, the NSF for funding through NSF grant AST-1450141, ``Collaborative Research: Accurate cosmology with strong gravitational lens time-delays''.
S.H.S. acknowledges support from the Max Planck Society through the Max Planck Research Group.  
K.C.W. is supported in part by an EACOA Fellowship awarded by the East Asia Core Observatories Association, which consists of the Academia Sinica Institute of Astronomy and Astrophysics, the National Astronomical Observatory of Japan, the National Astronomical Observatories of the Chinese Academy of Sciences, and the Korea Astronomy and Space Science Institute.
S.H. acknowledges support by the DFG cluster of excellence \lq{}Origin and Structure of the Universe\rq{} (\href{http://www.universe-cluster.de}{\texttt{www.universe-cluster.de}}).
C.E.R and C.D.F. were funded through the NSF grant AST-1312329,
``Collaborative Research: Accurate cosmology with strong gravitational
lens time-delays.
AJS acknowledges support by NASA through STSCI grant HST-GO-15320.
P.J.M. acknowledges support from the U.S.\ Department of Energy under contract number DE-AC02-76SF00515.
LVEK is supported in part through an NWO-VICI career grant (project number 639.043.308).

Based on observations collected at the European Organisation for Astronomical Research in the Southern Hemisphere
under ESO programme(s) 091.A-0642(A) (PI: Sluse), and 074.A-0302(A) (PI: Rix), 60.A-9306(A), 097.A-0454(A) (PI: Sluse), 090.A-0531(A) (PI. Fassnacht).
Based on observations obtained at the Gemini Observatory (PID: GS-2013A-Q-2, PI: Treu), which is operated by the Association of Universities for Research in Astronomy, Inc., under a cooperative agreement with the NSF on behalf of the Gemini partnership: the National Science Foundation (United States), the National Research Council (Canada), CONICYT (Chile), Ministerio de Ciencia, Tecnolog\'{i}a e Innovaci\'{o}n Productiva (Argentina), and Minist\'{e}rio da Ci\^{e}ncia, Tecnologia e Inova\c{c}\~{a}o (Brazil).

Funding for the DES Projects has been provided by the U.S. Department of Energy, the U.S. National Science Foundation, the Ministry of Science and Education of Spain, the Science and Technology Facilities Council of the United Kingdom, the Higher Education Funding Council for England, the National Center for Supercomputing Applications at the University of Illinois at Urbana-Champaign, the Kavli Institute of Cosmological Physics at the University of Chicago, the Center for Cosmology and Astro-Particle Physics at the Ohio State University, the Mitchell Institute for Fundamental Physics and Astronomy at Texas A\&M University, Financiadora de Estudos e Projetos, Funda{\c c}{\~a}o Carlos Chagas Filho de Amparo {\`a} Pesquisa do Estado do Rio de Janeiro, Conselho Nacional de Desenvolvimento Cient{\'i}fico e Tecnol{\'o}gico and the Minist{\'e}rio da Ci{\^e}ncia, Tecnologia e Inova{\c c}{\~a}o, the Deutsche Forschungsgemeinschaft and the Collaborating Institutions in the Dark Energy Survey. 

The Collaborating Institutions are Argonne National Laboratory, the University of California at Santa Cruz, the University of Cambridge, Centro de Investigaciones Energ{\'e}ticas, Medioambientales y Tecnol{\'o}gicas-Madrid, the University of Chicago, University College London, the DES-Brazil Consortium, the University of Edinburgh, the Eidgen{\"o}ssische Technische Hochschule (ETH) Z{\"u}rich, Fermi National Accelerator Laboratory, the University of Illinois at Urbana-Champaign, the Institut de Ci{\`e}ncies de l'Espai (IEEC/CSIC), the Institut de F{\'i}sica d'Altes Energies, Lawrence Berkeley National Laboratory, the Ludwig-Maximilians Universit{\"a}t M{\"u}nchen and the associated Excellence Cluster Universe, the University of Michigan, the National Optical Astronomy Observatory, the University of Nottingham, The Ohio State University, the University of Pennsylvania, the University of Portsmouth, SLAC National Accelerator Laboratory, Stanford University, the University of Sussex, Texas A\&M University, and the OzDES Membership Consortium.

Based in part on observations at Cerro Tololo Inter-American Observatory, National Optical Astronomy Observatory, which is operated by the Association of Universities for Research in Astronomy (AURA) under a cooperative agreement with the National Science Foundation.

The DES data management system is supported by the National Science Foundation under Grant Numbers AST-1138766 and AST-1536171. The DES participants from Spanish institutions are partially supported by MINECO under grants AYA2015-71825, ESP2015-66861, FPA2015-68048, SEV-2016-0588, SEV-2016-0597, and MDM-2015-0509, some of which include ERDF funds from the European Union. IFAE is partially funded by the CERCA program of the Generalitat de Catalunya. Research leading to these results has received funding from the European Research Council under the European Union's Seventh Framework Program (FP7/2007-2013) including ERC grant agreements 240672, 291329, and 306478. We  acknowledge support from the Brazilian Instituto Nacional de Ci\^encia e Tecnologia (INCT) e-Universe (CNPq grant 465376/2014-2).

This manuscript has been authored by Fermi Research Alliance, LLC under Contract No. DE-AC02-07CH11359 with the U.S. Department of Energy, Office of Science, Office of High Energy Physics. The United States Government retains and the publisher, by accepting the article for publication, acknowledges that the United States Government retains a non-exclusive, paid-up, irrevocable, world-wide license to publish or reproduce the published form of this manuscript, or allow others to do so, for United States Government purposes.

Based on observations made with the NASA/ESA Hubble Space Telescope, obtained at the Space Telescope Science Institute, which is operated by the Association of Universities for Research in Astronomy, Inc., under NASA contract NAS 5-26555. These observations are associated with program \#12889. Support for program \#12889 was provided by NASA through a grant from the Space Telescope Science Institute, which is operated by the Association of Universities for Research in Astronomy, Inc., under NASA contract NAS 5-26555.
This work is based in part on observations made with the Spitzer Space Telescope, which is operated by the Jet Propulsion Laboratory, California Institute of Technology under a contract with NASA.
This publication makes use of data products from the Two Micron All Sky Survey, which is a joint project of the University of Massachusetts and the Infrared Processing and Analysis Center/California Institute of Technology, funded by the National Aeronautics and Space Administration and the National Science Foundation.

This work made extensive use of TOPCAT \citep{Taylor2014}, and of Astropy, a community-developed core Python package for Astronomy \citep{Astropy2013}. 



\bibliographystyle{../common/mnras}
\bibliography{WFI2033_lensenv_tmp.bib}

\section*{AFFILIATIONS}

$^{1}$ STAR Institute, Quartier Agora - All\'ee du six Ao\^ut, 19c B-4000 Li\`ege, Belgium\\
$^{2}$ Subaru Telescope, National Astronomical Observatory of Japan, 650 N Aohoku Pl, Hilo, HI 96720\\
$^{3}$ Department of Physics, University of California, Davis, CA 95616, USA \\
$^{4}$ Kavli IPMU (WPI), UTIAS, The University of Tokyo, Kashiwa, Chiba 277-8583, Japan \\
$^{5}$ National Astronomical Observatory of Japan, 2-21-1 Osawa, Mitaka, Tokyo 181-8588, Japan\\
$^{6}$ Max Planck Institute for Astrophysics, Karl-Schwarzschild-Strasse 1, D-85741 Garching, Germany\\
$^{7}$ Institute of Astronomy and Astrophysics, Academia Sinica, P.O.~Box 23-141, Taipei 10617, Taiwan\\
$^{8}$Physik-Department, Technische Universit{\"a}t M{\"u}nchen, James-Franck-Strasse~1, 85748 Garching, Germany \\
$^9$ European Southern Observatory, Karl-Schwarzschild-Strasse 2, D-85748 Garching, Germany. \\
$^{10}$ Institute of Astronomy, University of Cambridge, Madingley Road, Cambridge CB3 0HA, UK \\
$^{11}$ CRAL, Observatoire de Lyon, Saint-Genis-Laval,  France \\
$^{12}$ Department of Physics and Astronomy, PAB, 430 Portola Plaza, Box951547, Los Angeles, CA 90095, USA \\
$^{13}$ Laboratoire d'Astrophysique, Ecole Polytechnique F{\'e}d{\'e}rale de Lausanne (EPFL), Observatoire de Sauverny, CH-1290 Versoix, Switzerland\\
$^{14}$ Institute of Cosmology and Gravitation, University of Portsmouth, Burnaby Rd, Portsmouth PO1 3FX, UK \\
$^{15}$ Exzellenzcluster Universe, Boltzmannstr. 2, 85748 Garching, Germany\\
$^{16}$ Kapteyn Astronomical Institute, University of Groningen, PO Box 800, NL-9700 AV Groningen, The Netherlands\\
$^{17}$ Kavli Institute for Particle Astrophysics \& Cosmology, P. O. Box 2450, Stanford University, Stanford, CA 94305, USA \\  
$^{18}$ Fermi National Accelerator Laboratory, P. O. Box 500, Batavia, IL 60510, USA \\
$^{19}$ Instituto de Fisica Teorica UAM/CSIC, Universidad Autonoma de Madrid, 28049 Madrid, Spain \\
$^{20}$ CNRS, UMR 7095, Institut d'Astrophysique de Paris, F-75014, Paris, France \\
$^{21}$ Sorbonne Universit\'es, UPMC Univ Paris 06, UMR 7095, Institut d'Astrophysique de Paris, F-75014, Paris, France \\
$^{22}$ Department of Physics \& Astronomy, University College London, Gower Street, London, WC1E 6BT, UK \\
$^{23}$ Leiden Observatory, Leiden University, Niels Bohrweg 2, 2333 CA Leiden, the Netherlands \\
$^{24}$ SLAC National Accelerator Laboratory, Menlo Park, CA 94025, USA \\
$^{25}$ Centro de Investigaciones Energ\'eticas, Medioambientales y Tecnol\'ogicas (CIEMAT), Madrid, Spain \\
$^{26}$ Laborat\'orio Interinstitucional de e-Astronomia - LIneA, Rua Gal. Jos\'e Cristino 77, Rio de Janeiro, RJ - 20921-400, Brazil \\
$^{27}$ Department of Astronomy, University of Illinois at Urbana-Champaign, 1002 W. Green Street, Urbana, IL 61801, USA \\
$^{28}$ National Center for Supercomputing Applications, 1205 West Clark St., Urbana, IL 61801, USA \\
$^{29}$ Institut de F\'{\i}sica d'Altes Energies (IFAE), The Barcelona Institute of Science and Technology, Campus UAB, 08193 Bellaterra (Barcelona) Spain \\
$^{30}$ Institut d'Estudis Espacials de Catalunya (IEEC), 08034 Barcelona, Spain \\
$^{31}$ Institute of Space Sciences (ICE, CSIC),  Campus UAB, Carrer de Can Magrans, s/n,  08193 Barcelona, Spain \\
$^{32}$ Observat\'orio Nacional, Rua Gal. Jos\'e Cristino 77, Rio de Janeiro, RJ - 20921-400, Brazil \\
$^{33}$ Department of Physics, IIT Hyderabad, Kandi, Telangana 502285, India \\
$^{34}$ Department of Astronomy, University of Michigan, Ann Arbor, MI 48109, USA \\
$^{35}$ Department of Physics, University of Michigan, Ann Arbor, MI 48109, USA \\
$^{36}$ Kavli Institute for Cosmological Physics, University of Chicago, Chicago, IL 60637, USA  \\
$^{37}$ California Institute of Technology, 1200 East California Blvd, MC 249-17, Pasadena, CA 91125, USA  \\
$^{38}$ Department of Physics, ETH Zurich, Wolfgang-Pauli-Strasse 16, CH-8093 Zurich, Switzerland \\
$^{39}$ Santa Cruz Institute for Particle Physics, Santa Cruz, CA 95064, USA \\
$^{40}$ Center for Cosmology and Astro-Particle Physics, The Ohio State University, Columbus, OH 43210, USA  \\
$^{41}$ Department of Physics, The Ohio State University, Columbus, OH 43210, USA \\
$^{42}$ Center for Astrophysics $\vert$ Harvard \& Smithsonian, 60 Garden Street, Cambridge, MA 02138, USA  \\
$^{43}$ Lawrence Berkeley National Laboratory, 1 Cyclotron Road, Berkeley, CA 94720, USA  \\
$^{44}$ Department of Astronomy/Steward Observatory, University of Arizona, 933 North Cherry Avenue, Tucson, AZ 85721-0065, USA  \\
$^{45}$ Australian Astronomical Optics, Macquarie University, North Ryde, NSW 2113, Australia  \\
$^{46}$ Departamento de F\'isica Matem\'atica, Instituto de F\'isica, Universidade de S\~ao Paulo, CP 66318, S\~ao Paulo, SP, 05314-970, Brazil  \\
$^{47}$ George P. and Cynthia Woods Mitchell Institute for Fundamental Physics and Astronomy, and Department of Physics and Astronomy, Texas A\&M University, College Station, TX 77843,  USA  \\
$^{48}$ Department of Astrophysical Sciences, Princeton University, Peyton Hall, Princeton, NJ 08544, USA \\
$^{49}$ Instituci\'o Catalana de Recerca i Estudis Avan\c{c}ats, E-08010 Barcelona, Spain  \\
$^{50}$ Institut de F\'{\i}sica d'Altes Energies (IFAE), The Barcelona Institute of Science and Technology, Campus UAB, 08193 Bellaterra (Barcelona) Spain  \\
$^{51}$ School of Physics and Astronomy, University of Southampton,  Southampton, SO17 1BJ, UK \\
$^{52}$ Brandeis University, Physics Department, 415 South Street, Waltham MA 02453  \\
$^{53}$ Instituto de F\'isica Gleb Wataghin, Universidade Estadual de Campinas, 13083-859, Campinas, SP, Brazil  \\
$^{54}$ Computer Science and Mathematics Division, Oak Ridge National Laboratory, Oak Ridge, TN 37831


\appendix

\section{Comparison with literature redshifts}
\label{Appendix:redshift}

\begin{figure}
	\includegraphics[width=\columnwidth]{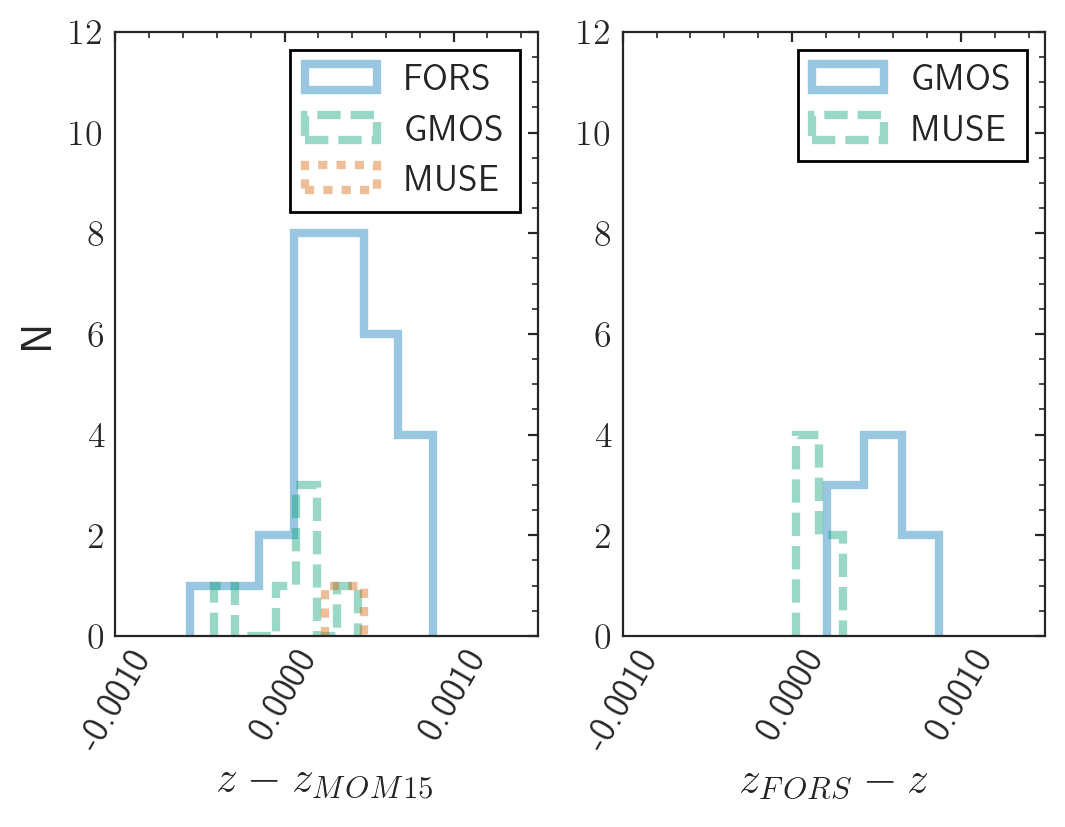}
	\caption{Left panel: distribution of the difference of redshifts  between our spectra obtained with 3 different instruments (FORS, GMOS, MUSE) and MOM15 (i.e. $\delta z = z_i - z_{\rm MOM15} $, where $i = [FORS, GMOS, MUSE]$). Right panel: Distribution of the difference of redshifts between FORS and GMOS / MUSE. }
	\label{fig:redshiftcompa}
\end{figure}

Independent redshift measurements are available for a small subsample of objects. On one hand, we have compared the 38 robust galaxy measurements present in our catalog and in MOM15 catalog\footnote{We only consider objects with the same redshift and with flags 3 and 4 in the MOM15, i.e. we exclude objects that are not new measurements from MOM15 but included in their catalog.}. The results (split by instrument), are shown in the left panel of Fig.~\ref{fig:redshiftcompa}. We find a median redshift offset with MOM15 of $\delta z = 3.6\times10^{-4} \pm 8.5\times10^{-6}$ (standard error on the mean; stde), compatible with what we observed for \HEofor. Since the width of the observed $\delta z$ distribution is compatible with the median formal error on the redshift, we conclude that, while GMOS and MOM15 redshifts are mutually compatible, the offset between FORS and MOM15 is real. On the other hand, we have compared  redshifs for targets in common in the FORS, GMOS and MUSE catalogs (right panel of Fig.~\ref{fig:redshiftcompa}). There are 9 galaxies with robust redshift measurements in common between the GMOS and FORS catalogs. The distribution of redshift differences between the 2 catalogs is centered on $\delta z = z_{FORS} - z_{GMOS} \sim 0.0005 \pm 0.00007$ (stde). There are 6 galaxies in common between the MUSE and the FORS samples, and we measure $\delta z = z_{MUSE} - z_{FORS} \sim -0.00009 \pm 0.00005$ (stde). Despite the smaller sample of galaxies, those results support the results found by comparing our results to MOM15, namely that FORS and MUSE redshifts are mutually compatible but offset by about $\delta z \sim 0.0004$ compared to MOM15 and GMOS.  

These comparisons suggest that there are unaccounted systematic errors between the catalogs, but it is difficult to trace their origin. This has no impact on our analysis provided that we apply a systematic correction to match redshifts of all the catalogs. Similarly to the approach followed for \HEofor, we have decided to correct the ESO-based data (i.e. MUSE and FORS). We apply for our analysis a systematic correction of  $\delta_z = -3.6\times10^{-4} $ to the FORS and MUSE redshifts. Such an offset corresponds to $\sim$ 1 FORS pixel $\sim$ 3.3\,\AA\, in our wavelength calibration (i.e. about five times the error on the wavelength calibration derived along the reduction). This translates into a velocity offset $\delta v \sim 120$\,\kms.  In addition, we have quadratically added an error $\sigma_{z}^{\rm{sys}}$ to the redshifts measured with different instruments. Based on the above comparisons, we have considered $\sigma_{z}^{\rm{sys}} = 0.0002$ for FORS, MUSE and GMOS, and $\sigma_{z}^{\rm{sys}} = 0.0001$ for MOM15 data. This systematic error accounts for the uncertainty on the offset between pairs of catalogs, and for a possible systematic error associated to the barycentric correction (our redshifts include a barycentric correction -smaller than the systematic offset $\delta z$- but we do not know if this is the case of MOM15 redshifts). Finally, we note that we do not account for the displacement of our Galaxy w.r.t. the Cosmic Microwave Background, which for \WFItwenty\, implies an additional correction $\delta z \sim 0.00053$ (that would be identical for all the catalogs). The impact of this correction on the cosmological analysis is yet negligible relative to other sources of uncertainties, and can otherwise be accounted for explicitly. 

In addition, there are also 4 objects for which our spectra do not support the redshift measurement published by \cite{Momcheva2015}. There is one object for which we propose to revise the redshift measurement. The discrepancy is small ($\delta z = 0.04$) but significant. Since there is a nearby, but fainter galaxy 3.0\arcsec away from the target, we cannot fully exclude that the redshifts correspond to 2 different objects, but this seems unlikely. For three other targets, we could not measure any redshift, while \cite{Momcheva2015} publish a robust measurement. Since the signal to noise ratio of our spectra is on average better, we suggest that there could be a potential mis-identification in \cite{Momcheva2015}. Therefore, we choose for this analysis to set the quality flag to 2 when there is no confirmation of the published redshift with our new spectrum, and to 1 in case of possible support. Table~\ref{tab:badredshifts} lists the coordinates and IDs of the galaxies with potentially discrepant redshifts. 

\begin{table*}
\caption{Objects with significantly different redshifts in Momcheva et al. (\citeyear{Momcheva2015}) and in our catalog. The last column comments on the reason of likely mis-identification. }
	\label{tab:badredshifts}
	\centering
	\begin{tabularx}{0.92\linewidth}{llllll}
		\toprule
		(RA,DEC) (deg) & ID-MOM & ID & $z_{\rm MOM}$ ($\sigma_{z}$) & $z$ ($\sigma_z$) & Note \\
		\midrule
		(308.417400, $-$47.41113)  	& 13198 & 354   & 0.3540 (3.0e-4)    & 0.39485 (9.7e-5)  & Unknown. Nearest galaxy is 3\arcsec. \\ 
		(308.390047, $-$47.39731)     & 11787 & 487   & 0.4997 (3.0e-4)    & -                 & Unknown. Flagged changed to 2 \\  
		(308.388206, $-$47.39150)     & 12266 & 1044  & 0.9701 (3.0e-4)    & -                 & Unknown. Flagged changed to 2 \\  
		(308.4031, $-$47.34367)       & 12814 & 607   & 0.7514 (2.3e-4)    & -                 & Unknown. Flagged changed to 1 \\ 
		\bottomrule
		
	\end{tabularx}

\end{table*}


\section{Groups found when using a luminosity weighting scheme}
\label{sec:groups_lweight}

As a comparison, we applied our group detection algorithm using a luminosity weighting scheme to calculate the group centroid. Table~\ref{tab:groups_lweight} displays the properties of the groups identified following this methodology. Groups at the same redshift and with similar velocity-dispersion as the fiducial ones are identified, except the group \texttt{a0-1} (Table~\ref{tab:groups_lweight}). On the other hand, the group candidates identified with that weighting scheme seem to be less physically plausible, in particular because they often host fewer galaxies within one virial radius than in the fiducial case. This led to the flagging of the groups \texttt{b5} ($\bar{z}_{\rm {group}}=0.3059$) and \texttt{a0-2} ($\bar{z}_{\rm {group}}=0.3867$) as spurious group candidates. The group centroid is compatible with the one reported in Table~\ref{tab:groups} within 3~$\sigma$. The group centroid is generally found to be closer to a group galaxy in the fiducial case. In addition, the brightest galaxy is never found to be the closest to the centroid, even when luminosity weighting is used. 

\begin{table*}
	\caption{Properties of the groups identified in the FOV of \WFItwenty. Same as Table~\ref{tab:groups} but considering a luminosity weighted centroid. The compact group a2-1 is not identified when we follow that methodology. The columns are the group redshift, the number of spectroscopically identified galaxies in the group, the group intrinsic velocity-dispersion (rounded to the nearest 10\,\kms) and 1$\sigma$ standard deviation from bootstrap, the group centroid, bootstrap error on the centroid, projected distance of the centroid to the lens, median flexion shift $\log(\Delta_3 x (arcsec))$ and 1$\sigma$ standard deviation from bootstrapping (Sect.~\ref{sec:model}). The last column indicates for which field a peak of more than 5 galaxies is detected in redshift space. The properties we display correspond to the FOV marked in bold. }
	\label{tab:groups_lweight}
	\centering
	\begin{minipage}{\linewidth}
		\centering
	\begin{tabular*}{0.9\linewidth}{llrcccccc}
		\hline
		ID & $\bar{z}_{\rm group}$ & N & $\sigma_{\rm{int}}$ $\pm$ err& $\alpha_{\rm ctr}$, $\delta_{\rm ctr}$ & err($\alpha_{\rm ctr}$, $\delta_{\rm ctr}$) &  $\Delta\theta$ & $\log(\Delta_3 x) \pm $ err & FOV  \\
		& & & \kms & deg &  arcsec &  arcsec   & $\log(\rm{arcsec})$ & arcmin \\
\hline

b5 & 0.3059$^\dagger$ & 10 & 490 (90) & 308.50539418, $-$47.37161310 &  251.0, 145.1  & 212.7  & $-$6.07 $\pm$ 0.63 & 900 \\ 
a0-2 & 0.3867$^\dagger$ & 8 & 100 (20) & 308.44570176, $-$47.36695669 &  87.9, 20.4  & 113.8 & $-$7.68 $\pm$ 0.51 & 360 \\ 
a0-3 & 0.3999$^\dagger, \ddagger$ & 12 & 380 (70) & 308.34270600, $-$47.33671275 &  87.6, 77.7  & 292.5 & $-$7.04 $\pm$ 0.44 & 360 \\
a2 & 0.4436 & 6 & 150 (40) & 308.47252846, $-$47.39111478 &  19.3, 25.7  & 115.5 & $-$7.18 $\pm$ 0.46 &  360 \\
a3 & 0.4956 & 13 & 520 (110) & 308.42814292, $-$47.38372192 &  67.3, 66.6  & 42.8 & $-$4.43 $\pm$ 0.95 & 360 \\ 
a5 & 0.6588 & 22 & 500 (90) &  308.44512558, $-$47.37258509 &  36.0, 20.2 & 95.2 & $-$4.87 $\pm$ 0.43& {\bf {360}}, 900 \\ 
a8 & 0.6796$^\ddagger$ & 11 & 610 (200) & 308.42531059, $-$47.39318538 &  71.5, 25.2  & 8.3 & $-$3.71 $\pm$ 1.14 & {\bf {360}}, 900  \\ 
a9 & 0.6889 & 4 & 190 (90) & 308.41390309, $-$47.41358372 &  22.9, 15.3  & 71.1 & $-$6.37 $\pm$ 2.20 & 360 \\


\hline
\end{tabular*}
\begin{flushleft}
{\small 
	{\bf Note:} ${\dagger}$ Likely spurious. $\ddagger$ No luminosity weighting due to some group members missing magnitude measurement. $\star$ Results from the break down of a larger multi-modal group candidate. \\
}
\end{flushleft}

\end{minipage}

\end{table*}

\section{Masking for velocity-dispersion measurement}
\label{Appendix:skymask}

Two different masking schemes have been used to measure the velocity-dispersion of $G$. A first mask consists of regions identified as being susceptible to residual sky artefacts, as well as main telluric lines. The second mask contains the same features as the first mask but wavelengths bluer than 6600\,\AA\, are also masked out, as those regions are the most susceptible to uncertainties due to sky extraction. For G2, a third mask, containing only the known telluric absorption and the red edge of the spectrum ($\lambda > 9100\,$\AA) has also been used.  

We list in Table~\ref{tab:mask} the spectral range excluded for the velocity-dispersion measurements. This list contains known telluric absorption, as well as spectral ranges for which the data suggested sky variability larger than expectation based on the variance frame. To identify those regions, we proceeded as follows:  we extract a spectrum in an aperture of the same size as the aperture used for the galaxies, for 10 different empty regions of the FOV. We then compare the observed variation of the spectrum, as a function of wavelength, to the median variance estimated based on the variance spectrum. We have considered that regions (limited to maximum 50 consecutive pixels) that deviate by more than 3 times the median standard deviation are the most susceptible to be poorly sky subtracted, and therefore are masked out. Similar regions are found for SV and P97 data, suggesting that those regions need effectively to be masked in addition to the spectral bands containing telluric lines. In addition to those regions, we have also added the reddest edge of the spectrum whose absolute level varies by more than 100\% between individual data-cubes in the SV data. 

\begin{table}
	\caption{Sky bands masked out for the velocity-dispersion measurements. $\lambda_1$ and $\lambda_2$ correspond respectively to the beginning and end of the wavelength band. The third column indicates if the band is a region known to be affected by telluric absorption or identified based on the study of the sky variability. For clarity, we have not merged overlapping masked regions.}
	\label{tab:mask}
\begin{tabular}{ll|r}
		$\lambda_1$ (\AA) & $\lambda_2$ (\AA) & Note \\
		\hline
		4750 & 4780 & Telluric \\ 
		5573 & 5584 & Variable \\
		5885 & 5960 & Telluric \\ 
		6295 & 6307 & Variable \\ 
		6359 & 6369 & Variable \\ 
		7716 & 7782 & Variable \\ 
		7791 & 7857 & Variable \\ 
		7800 & 7920 & Telluric \\ 
		7910 & 7968 & Variable \\ 
		7990 & 8030 & Variable \\ 
		8341 & 8405 & Variable \\ 
		8426 & 8471 & Variable \\ 
		8758 & 8790 & Telluric \\ 
		8762 & 8829 & Variable \\ 
		8820 & 8850 & Telluric \\
		8826 & 8892 & Variable \\ 
		8885 & 8895 & Telluric \\ 
		8902 & 8965 & Variable \\ 
		8910 & 8922 & Telluric \\
		8997 & 9056 & Variable \\  
		9100 & 9344 & Red edge\\
		
	\end{tabular}
\end{table}

\onecolumn

\begin{landscape}
\scriptsize
\begin{longtable}{ccccccccccccccc}
\caption{Photometric properties of the $i\leq23$ galaxies inside $120\arcsec$ of \WFItwenty } \\ 
  \hline 
RA  & DEC & $u$ & $g$ & $r$ & $i$ & $z$ & $Y$ & $J$ & $H$ & $K$ & $3.6\mu m$ & $4.5\mu m$ & $5.7\mu m$ & $7.9\mu m$ \\ 
\hline
\input{table_phot.tex}
\end{longtable}
\vspace*{-7.0mm}
{\normalsize We report $\mathrm{MAG\_ISO}$ magnitudes measured with \texttt{\texttt{SExtractor}}, but corrected to total magnitudes using $\mathrm{MAG\_AUTO}_i - \mathrm{MAG\_ISO}_i$, with detections in the combined $ir$-bands. Reported magnitudes are corrected for galactic extinction following \citet{Schlegel1998}. For the IRAC channels, when computing the physical quantities in Table \ref{tab:zmstar}, additional uncertainties were added from the \texttt{EAzY} template error function (not included in this table).}
\end{landscape}

\begin{longtable}{ccccc|ccccc}
\caption{Measured redshifts and stellar masses of the galaxies in Table \ref{tab:phot}} \\ 
  \hline 
RA  & DEC & sep & zspec/\texttt{EAzY} & $\log M_\star$ & RA  & DEC & sep & zspec/\texttt{EAzY} & $\log M_\star$ \\
 \hline
\input{table_zMstar.tex}
\end{longtable}
\vspace*{-7.0mm}
{\normalsize Where spectroscopic measurements (marked with null error bars) are not available, photometric redshift values (estimated with \texttt{EAzY}) correspond to the peak of the probability distributions, and logarithmic mass values (incorporating the IRAC photometry) correspond to the medians of the probability distributions estimated with \texttt{\texttt{Le PHARE}}, unless only the $\mathrm{MASS\_BEST}$ value was successfully computed by \texttt{\texttt{Le PHARE}}. Error bars mark the enclosed 68\% confidence regions.}


\bsp	
\label{lastpage}
\end{document}

%% file: table_phot.tex
308.42101 & $-47.36379$ & $23.42 \pm 0.07$ & $23.65 \pm 0.13$ & $23.01 \pm 0.09$ & $22.44 \pm 0.09$ & $21.85 \pm 0.15$ & $21.76 \pm 0.21$ & $21.34 \pm 0.16$ & $21.06 \pm 0.17$ & $20.83 \pm 0.11$ & $20.96 \pm 0.01$ & $21.11 \pm 0.02$ & $21.86 \pm 0.18$ & $20.68 \pm 0.13$ \\
308.41123 & $-47.36396$ & $22.30 \pm 0.04$ & $21.38 \pm 0.02$ & $20.10 \pm 0.01$ & $19.55 \pm 0.01$ & $19.30 \pm 0.02$ & $19.05 \pm 0.04$ & $18.83 \pm 0.03$ & $18.44 \pm 0.03$ & $18.18 \pm 0.02$ & $18.88 \pm 0.00$ & $18.96 \pm 0.00$ & $19.40 \pm 0.02$ & $18.73 \pm 0.02$ \\
308.44054 & $-47.36504$ & $25.05 \pm 0.15$ & $24.53 \pm 0.21$ & $22.99 \pm 0.06$ & $21.76 \pm 0.03$ & $21.76 \pm 0.09$ & $21.30 \pm 0.14$ & $20.85 \pm 0.07$ & $20.39 \pm 0.07$ & $20.17 \pm 0.06$ & $20.03 \pm 0.01$ & $20.49 \pm 0.01$ & $20.55 \pm 0.04$ & $21.57 \pm 0.22$ \\
308.44739 & $-47.36605$ & $24.09 \pm 0.09$ & $24.09 \pm 0.16$ & $23.39 \pm 0.09$ & $22.83 \pm 0.09$ & $22.47 \pm 0.17$ & $-$ & $22.34 \pm 0.20$ & $-$ & $-$ & $22.17 \pm 0.03$ & $22.72 \pm 0.08$ & $-$ & $-$ \\
308.43790 & $-47.36645$ & $23.43 \pm 0.07$ & $23.00 \pm 0.07$ & $22.42 \pm 0.05$ & $22.18 \pm 0.06$ & $22.26 \pm 0.13$ & $-$ & $22.38 \pm 0.18$ & $21.43 \pm 0.21$ & $-$ & $22.24 \pm 0.04$ & $22.30 \pm 0.06$ & $-$ & $-$ \\
308.42581 & $-47.36886$ & $22.61 \pm 0.05$ & $22.36 \pm 0.05$ & $21.46 \pm 0.03$ & $20.84 \pm 0.03$ & $20.68 \pm 0.08$ & $20.39 \pm 0.10$ & $20.65 \pm 0.11$ & $20.12 \pm 0.11$ & $20.10 \pm 0.11$ & $20.03 \pm 0.01$ & $20.49 \pm 0.01$ & $20.47 \pm 0.05$ & $20.99 \pm 0.15$ \\

%% file: table_zMstar.tex
308.42101 & $-47.36379$ & 113.9 & $0.891_{-0.124}^{+0.129}$ & $10.40_{-0.10}^{+0.08}$ & 308.41123 & $-47.36396$ & 117.9 & $0.388_{-0.000}^{+0.000}$ & $10.66_{-0.04}^{+0.03}$\\
308.44054 & $-47.36504$ & 115.0 & $0.736_{-0.054}^{+0.051}$ & $10.45_{-0.04}^{+0.04}$ & 308.44739 & $-47.36605$ & 118.2 & $0.724_{-0.154}^{+0.161}$ & $9.50_{-0.50}^{+0.46}$\\
308.43790 & $-47.36645$ & 108.2 & $0.433_{-0.148}^{+0.127}$ & $7.99_{-0.21}^{+0.21}$ & 308.42581 & $-47.36886$ & 95.1 & $0.651_{-0.064}^{+0.065}$ & $10.18_{-0.06}^{+0.05}$\\
308.42481 & $-47.36999$ & 91.1 & $0.387_{-0.000}^{+0.000}$ & $9.71_{-0.07}^{+0.06}$ & 308.40978 & $-47.36709$ & 108.3 & $0.228_{-0.067}^{+0.068}$ & $8.64_{-0.62}^{+0.54}$\\
308.44013 & $-47.36891$ & 101.6 & $0.781_{-0.127}^{+0.146}$ & $9.89_{-0.47}^{+0.43}$ & 308.42281 & $-47.36942$ & 93.3 & $0.642_{-0.087}^{+0.088}$ & $9.19_{-0.66}^{+0.58}$\\
308.43034 & $-47.36987$ & 92.3 & $0.611_{-0.114}^{+0.111}$ & $9.07_{-0.52}^{+0.65}$ & 308.43807 & $-47.37058$ & 94.2 & $0.537_{-0.139}^{+0.126}$ & $9.42_{-0.09}^{+0.08}$\\